# A flexible method for estimating luminosity functions via Kernel Density Estimation - II. Generalization and Python implementation

Zunli Yuan,[1,2] Xibin Zhang,[3] Jiancheng Wang,[4,5,6] Xiangming Cheng,[4,5,6] and Wenjie Wang[1,2]

[1]*Department of Physics, School of Physics and Electronics, Hunan Normal University, Changsha 410081, China*
[2]*Key Laboratory of Low Dimensional Quantum Structures and Quantum Control, Hunan Normal University, Changsha 410081, China*
[3]*Department of Econometrics and Business Statistics, Monash University, Australia*
[4]*Yunnan Observatories, Chinese Academy of Sciences, Kunming 650216, China*
[5]*Key Laboratory for the Structure and Evolution of Celestial Objects, Chinese Academy of Sciences, Kunming 650216, China*
[6]*University of Chinese Academy of Sciences, Beijing 100049,China*

## ABSTRACT

We propose a generalization of our previous KDE (kernel density estimation) method for estimating luminosity functions (LFs). This new upgrade further extend the application scope of our KDE method, making it a very flexible approach which is suitable to deal with most of bivariate LF calculation problems. From the mathematical point of view, usually the LF calculation can be abstracted as a density estimation problem in the bounded domain of $\{Z_1 < z < Z_2, \ L > f_{\lim}(z)\}$. We use the transformation-reflection KDE method ($\hat{\phi}$) to solve the problem, and introduce an approximate method ($\hat{\phi}_1$) based on one-dimensional KDE to deal with the small sample size case. In practical applications, the different versions of LF estimators can be flexibly chosen according to the Kolmogorov-Smirnov test criterion. Based on 200 simulated samples, we find that for both cases of dividing or not dividing redshift bins, especially for the latter, our method performs significantly better than the traditional binning method $\hat{\phi}_{\rm bin}$. Moreover, with the increase of sample size $n$, our LF estimator converges to the true LF remarkably faster than $\hat{\phi}_{\rm bin}$. To implement our method, we have developed a public, open-source Python Toolkit, called `kdeLF`. With the support of `kdeLF`, our KDE method is expected to be a competitive alternative to existing nonparametric estimators, due to its high accuracy and excellent stability. `kdeLF` is available at http://github.com/yuanzunli/kdeLF with extensive documentation available at http://kdelf.readthedocs.org/en/latest .

## 1. INTRODUCTION

Luminosity function (LF) is a very fundamental statistic, which provides one of the most important tools to probe the distribution and evolution of galaxies and active galactic nuclei (AGNs) over cosmic time. Accurately measuring the LF of extragalactic objects has long been an important pursuit in modern astronomy. However, this is difficult in observational cosmology since the presence of observational selection effects (e.g., due to detection thresholds in flux density, apparent magnitude, color, surface brightness, etc.) can make any given galaxy survey incomplete and thus introduce biases into the LF estimate (e.g., Marchetti et al. 2016). Numerous statistical approaches have been developed to overncome this limit. These mainly include parametric techniques (e.g., Sandage et al. 1979; Marshall et al. 1983; Jarvis & Rawlings 2000), and nonparametric methods such as the binned methods (e.g., Schmidt 1968; Avni & Bahcall 1980; Page & Carrera 2000, denoted as $\hat{\phi}_{\rm bin}$), the Lynden-Bell (1971) $C^-$ estimator and its extended versions (e.g., Efron & Petrosian 1992; Caditz & Petrosian 1993; Singal et al. 2014).

Corresponding author: Zunli Yuan, Xibin Zhang
yzl@hunnu.edu.cn, xibin.zhang@monash.edu

The parametric technique itself has been very mature, and the main focus is on chosing appropriate physical models. Nonparametric estimators are advantageous in cases where either there is difficult to find a proper parametric physical model or there is a desire to validate a parametric model (e.g., Schafer 2007). They still have great room for development. On one hand, the traditional binned methods and $C^-$ estimator have insurmountable disadvantages (Yuan & Wang 2013; Kelly et al. 2008). On the other hand, some newly developed methods (e.g., Schafer 2007; Kelly et al. 2008; Takeuchi 2010) had hardly shaken the dominant position of traditional estimators, either due to issues of application scope or other factors like computational complexity.

From the mathematical point of view, LF calculation is a typical density estimation problem in the bounded domain. The mathematics behind $\hat{\phi}_{\rm bin}$ is the histogram, the oldest and also the simplest density estimator. The $C^-$ estimator resorts to empirical cumulative distributions to avoid the binning of data. The more recent LF methods adopted some more advanced mathematics such as the local likelihood density estimation (Schafer 2007), the copula method (Takeuchi 2010), the $k$th-nearest-neighbor method (Caditz 2018), and the mixture of Gaussians model in a Bayesian perspective (Kelly et al. 2008), etc.

Kernel density estimation (KDE) is a well-established nonparametric approach to estimate continuous density functions either in univariate or multivariate cases. After years of development and continuous renewal, KDE has become the most popular method for estimation, interpolation, and visualization of probability density functions (e.g., Silverman 1986; Botev et al. 2010; Zhang et al. 2014; Chen 2017; Davies & Baddeley 2018; Gramacki 2018). Caditz & Petrosian (1993) was probably the first to introduce the concept of KDE to LF calculation, while they mainly utilized KDE to improve the $C^-$ estimator. In the following years, the idea of KDE did not form a mature solutions for LF measurement. This is due to either the limitation of computing power or lack of data in earlier years. Both the factors restricted the advantages of KDE.

Yuan et al. (2020, hereafter Paper I) recently proposed a new method for estimating LFs in the framework of KDE. The new method, proved to be significantly more accurate and stable than $\hat{\phi}_{\rm bin}$ by simulation, has the advantages of both parametric and nonparametric methods. Nevertheless, the method was mainly devised to treat problems where the sample should have a coverage of redshift as broad as possible. In practice one is usually interested in the LF in some redshift interval $(Z_1, Z_2)$, e.g., estimating the local LF of extragalactic sources (e.g., Condon et al. 2002; Mauch & Sadler 2007; Sadler et al. 2014; Lofthouse et al. 2018; Symeonidis & Page 2019), or measuring the high-redshift LFs of galaxies (e.g., Bhatawdekar et al. 2019; Bowler et al. 2020), AGNs (e.g., Matsuoka et al. 2018; Wang et al. 2019; Adams et al. 2020), and Ly$\alpha$ emitters (e.g., Zheng et al. 2017; Konno et al. 2018; Hu et al. 2019). In this paper, we further develop the KDE method to suit more common application scenarios. We have also developed `kdeLF`: a public, open-source Python Toolkit. This code is designed to standardize the calculations of our method into a coherent, well-documented user interface. The efficiency and generality of the algorithm makes it capable of treating most of the LF problems for AGNs and galaxies.

Throughout the paper, we adopt a Lambda Cold Dark Matter cosmology with the parameters $\Omega_m = 0.30$, $\Omega_\Lambda = 0.70$, and $H_0 = 70$ km s$^{-1}$ Mpc$^{-1}$.

## 2. METHODOLOGY

### 2.1. *Kernel density estimation*

Suppose we observe $n$ points in a 2-dimensional (2D) space, $X = \{X_1, X_2, \cdots, X_n\}$. Assuming X arises from an unknown probability density function $f(\mathbf{x})$, the goal of kernel density estimation (KDE) is to estimate $f$. The classical fixed-bandwidth kernel estimate of $f$ given X is written as

$$\hat{f}(\mathbf{x}) = \frac{1}{nh_1h_2} \sum_{j=1}^{n} K\big(\frac{x_1 - X_{j,1}}{h_1}, \frac{x_2 - X_{j,2}}{h_2}\big), \tag{1}$$

where $X_{j,1}$ and $X_{j,2}$ are the $j$th observations of the first and second components of the vector $X$, and $n$ is the sample size. In this equation $K(\cdot, \cdot)$ is a bivariate kernel function (a non-negative probability density function that integrates to one), and $h_1$ and $h_2$ are smoothing parameters also called the bandwidths. It is well known that the choice of kernel function is a secondary matter in comparison with selection of the bandwidth (e.g., Wand & Jones 1995; Botev et al. 2010; Chen 2017; Gramacki 2018). The normal kernel is often used, i.e.,

$$K(x_1, x_2) = \frac{1}{2\pi} \exp(-\frac{1}{2}(x_1^2 + x_2^2)). \tag{2}$$

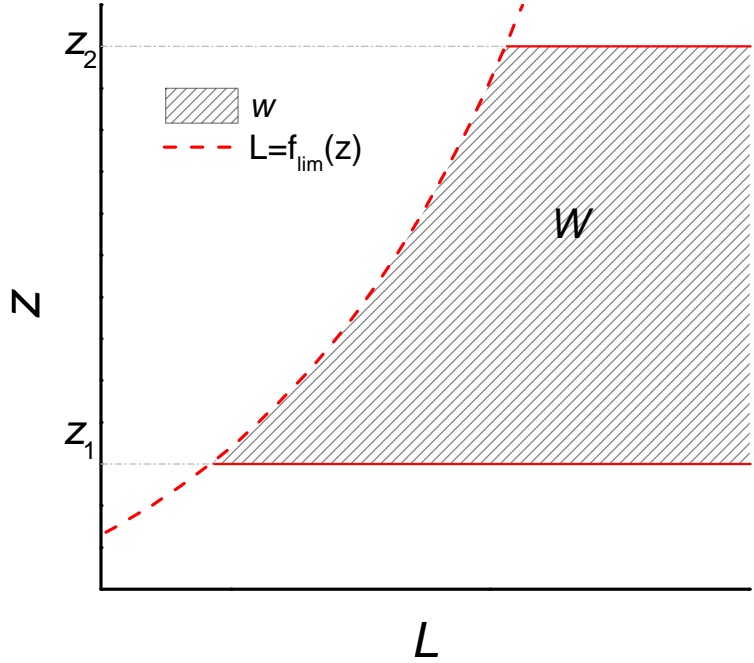

**Figure 1.** L-z plane showing the domain of data set.

### 2.2. Estimating the LF via KDE

The differential LF of a sample of objects is defined as the number of objects per unit comoving volume per unit luminosity interval,

$$\phi(z, \mathcal{L}) = \frac{d^2 N}{dV\, d\mathcal{L}}, \tag{3}$$

where $z$ denotes redshift and $\mathcal{L}$ denotes the luminosity. Due to the typically large span of the luminosities, it is often defined in terms of $\log \mathcal{L}$,

$$\phi(z, L) = \frac{d^2 N}{dV\, dL}, \tag{4}$$

where $L \equiv \log \mathcal{L}$ denotes the logarithm of luminosity. In practice, the main challenge is that only very limited objects are observed and one have to estimate the LF given truncated data. Figure 1 depicts the typical $L - z$ plane of a truncated data set. The domain of data is $W$ defined by $\{Z_1 < z < Z_2, L > f_{\text{lim}}(z)\}$; $W$ is referred to as the survey region (Paper I) or study window (e.g., Davies et al. 2018); $f_{\text{lim}}(z)$ is the truncation boundary of sample. For the common flux-limited samples, $f_{\text{lim}}(z)$ is given by

$$f_{\text{lim}}(z) = 4\pi d_L^2(z)(1/K(z))F_{\text{lim}}, \tag{5}$$

where $d_L(z)$ is the luminosity distance, $F_{\text{lim}}$ is the survey flux limit, and $K(z)$ represents the $K$-correction.

The case described in Figure 1 is general to nearly all kinds of LF probleams. If $Z_1 \simeq 0$ and $Z_2 \gg Z_1$, it is the case studied by Paper I; Else if the interval of $(Z_1, Z_2)$ is relatively small, this corresponds to artificial or forced division of redshift bins as done in many practices. In addition, $f_{\text{lim}}(z)$ can take any complicated form (e.g., Richards et al. 2006), not just the one given in Equation (5). The purpose of this work is to give an estimate to the LF $\phi$ over the survey region $W$. Following Paper I, $\phi$ is related to the probability distribution of $(z, L)$ by

$$p(z, L) = \frac{\Omega}{n} \phi(z, L) \frac{dV}{dz}, \tag{6}$$

where $\Omega$ is the solid angle subtended by the survey, and $dV/dz$ is the differential comoving volume per unit solid angle (Hogg 1999); $n$ is the number of objects observed within $(Z_1, Z_2)$, and hereafter it is referred to as "sample size". The domain of $p(z, L)$ is also $W$. Seeing that $W$ is a bounded area, direct estimate of $p(z, L)$ via KDE based on the data set $\{(z_1, L_1), (z_2, L_2), ...(z_n, L_n)\}$ is difficult. Serious bias can arise at the boundaries and near them, known as boundary effects or boundary bias (e.g., Müller & Stadtmüller 1999). The following transformation is made:

$$x = \ln\left(\frac{z - Z_1}{Z_2 - z}\right), \text{ and }, y = L - f_{\lim}(z). \tag{7}$$

The determinant of Jacobian matrix for the transformation is

$$\det(\mathbf{J}) = |\mathbf{J}| = \frac{Z_2 - Z_1}{(z - Z_1)(Z_2 - z)}. \tag{8}$$

Following the transformation-reflection method of Paper I, we first transform all pairs $(z_i, L_i)$ to $(x_i, y_i)$ according to Equation (7), then add the missing "probability mass" (e.g., Gramacki 2018) represented by the data set $(x_1, -y_1), (x_2, -y_2), ..., (x_n, -y_n)$. Figure 2 provides a toy example showing how a data set with the bounded domain of $\{Z_1 < z < Z_2, L > f_{\lim}(z)\}$ (top panel) will look like after transformation-reflection (bottom panel). Therefore, the purpose of transformation-reflection is to achieve an unbounded data set to which the KDE estimator can be deployed. The density of (x, y), denoted as[1] $\hat{f}(x, y)$, can be estimated by

$$\hat{f}(x, y) = \frac{2}{2nh_1 h_2} \sum_{j=1}^{n} \left( K(\frac{x-x_j}{h_1}, \frac{y-y_j}{h_2}) + K(\frac{x-x_j}{h_1}, \frac{y+y_j}{h_2}) \right). \tag{9}$$

Then the density of original data set, $(z, L)$, is

$$\hat{p}(z, L|h_1, h_2) = \hat{f}(x, y)\det(\mathbf{J}). \tag{10}$$

where the bandwidths, $h_1$ and $h_2$, are the only unknown parameters that need to be determined.

### 2.3. *Optimal Bandwidth Selection*

Bandwidth selection is the most important issue for using KDE, because the accuracy of KDE is mainly determined by the chosen bandwidths (e.g., Gramacki 2018). To find the optimal bandwidth, we use the likelihood cross-validation (LCV) criterion of Paper I, but make slight modifications described as follows.

$$S = -2\sum_{i}^{n} \ln[\hat{p}_{(-i)}(z_i, L_i|h_1, h_2)] + 2n \int_{Z_1}^{Z_2}\int_{f_{\lim}(z)}^{L_{\max}} \hat{p}(z, L|h_1, h_2) dz dL. \tag{11}$$

$\hat{p}_{-i}(z_i, L_i|h_1, h_2)$ is calculated by

$$\hat{p}_{(-i)}(z_i, L_i|h_1, h_2) = \frac{(Z_2-Z_1)\hat{f}_{(-i)}(x_i, y_i)}{(z-Z_1)(Z_2-z)}, \tag{12}$$

where

$$\hat{f}_{(-i)}(x_i, y_i) = \frac{2}{(2n - \eta_i)h_1 h_2}\left(\sum_{j \in J_i} K(\frac{x_i-x_j}{h_1}, \frac{y_i-y_j}{h_2}) + \sum_{j \in J_i'} K(\frac{x_i-x_j}{h_1}, \frac{y_i+y_j}{h_2})\right), \tag{13}$$

where $J_i = \{j : x_i \neq x_j \text{ and } y_i \neq y_j, \text{for } j = 1, 2, ..., n\}$, and $J_i' = \{j : x_i \neq x_j, \text{for } j = 1, 2, ..., n\}$, for i=1,2,...n. Note that $\hat{f}_{(-i)}(x_i, y_i)$ is slightly different from the *leave-one-out* estimator of Paper I (see their Equation 26). The purpose of the conventional leave-one-out estimator is to exclude an observation that makes the argument of the kernel function zero. Otherwise, the resulting kernel density estimator would contain an unwanted term $K(0/h_1, 0/h_2)$, and

---

[1] In the following text, the two expressions $\hat{f}(x, y)$ and $\hat{f}(x, y|h_1, h_2)$ are equivalent. The latter is to emphasize that $h_1$ and $h_2$ are parameters. Similarly for others, e.g., $\hat{f}_a(x, y)$ and $\hat{f}_a(x, y|h_{10}, h_{20}, \beta)$, etc.

Estimating luminosity functions via KDE 5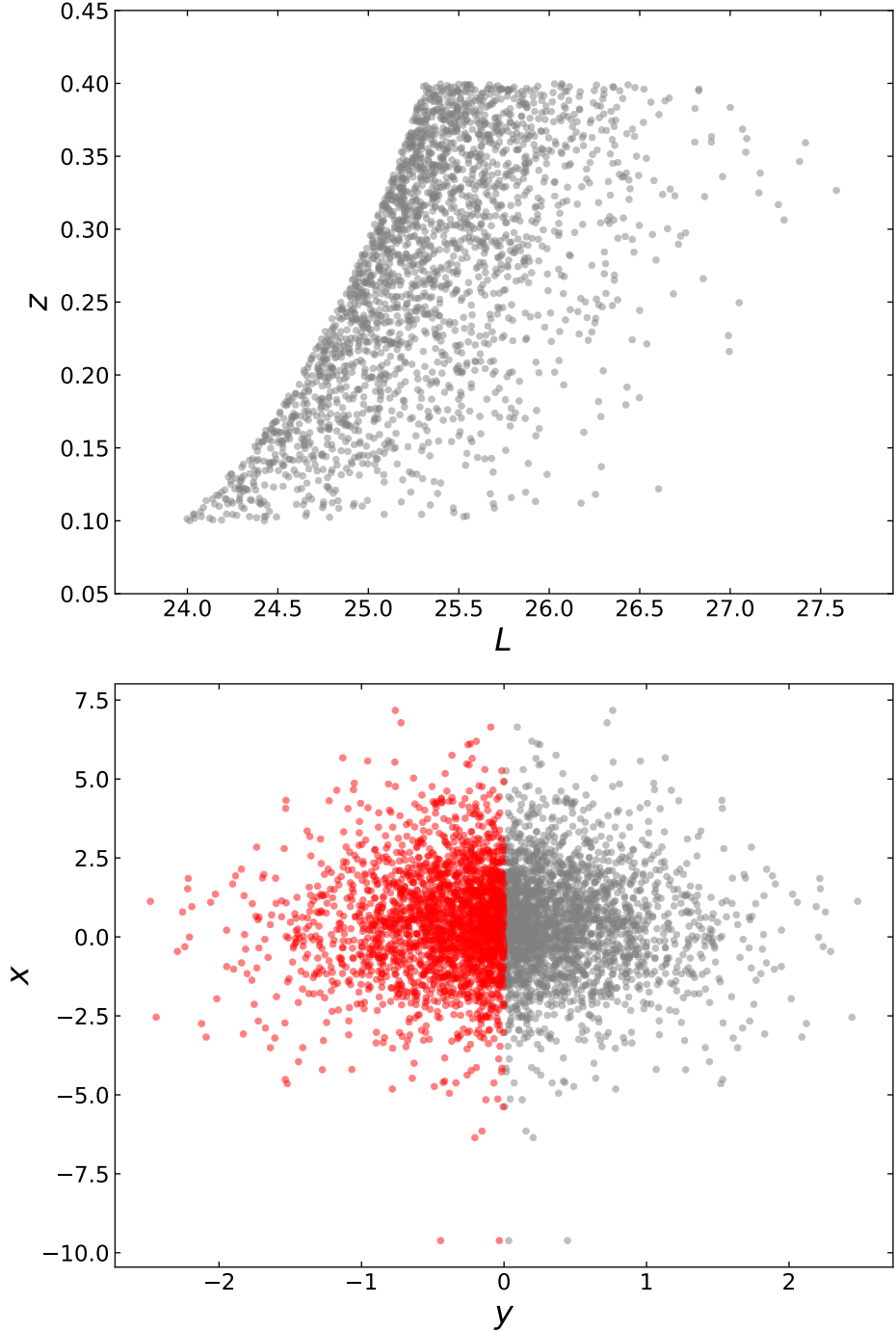

**Figure 2.** Top panel: A example data set with the bounded domain of $\{Z_1 < z < Z_2,\ L > f_{\text{lim}}(z)\}$. Bottom panel: It shows how the data in top panel look like after transformation (light gray points) by Equation (7), and adding the reflection points (red points).

minimisation of the cross-validated likelihood would approach infinity as $h_1$ and $h_2$ tend to zero. However, we find that it is not enough to simply leave one identical observation out, because there are repeated observations (mainly the redshift). This will lead to the unwanted terms of $0/h_1$ or $0/h_2$ even when $j \neq i$. Thus, at any point $(x_i, y_i)$, we propose excluding all the $j$th terms that make $x_i = x_j$ or $y_i = y_j$ from the first summation, and $x_i = x_j$ from the second summation in Equation (13), and $\eta_i$ denotes the total number of terms that are excluded. Similar ideas can be found in Zhang et al. (2014). We refer $\hat{f}_{(-i)}(x_i, y_i)$ as the *leave-more-out* estimator.

In Equation (11), $L_{\max}$ is the higher luminosity limit of the survey region $W$. Because our likelihood function was derived from Marshall et al. (1983), in paper I, we let $L_{\max}$ take a value slightly larger than $\max(L_i)$ according to the conventions in the literature (e.g., Cara & Lister 2008; Ajello et al. 2012). But we should note that the basis for the conventional value of $L_{\max}$ is ambiguous. Can we make it take a much larger value? Obviously, the answer is yes, as usually a sample does not have an observed maximum luminosity limit. If we let $L_{\max} \gg \max(L_i)$, the double integration in Equation (11) $\simeq 1$, and is independent of the values of $h_1$ and $h_2$. So we may drop the double integration term and obtain a more simplified LCV criterion given by

$$S_0 = -2 \sum_i^n \ln[\hat{p}_{(-i)}(z_i, L_i|h_1, h_2)] \tag{14}$$

The optimal bandwidths $h_1, h_2$ can be obtained by numerically minimizing the objective function $S$ or $S_0$. Note that $S$ and $S_0$ are the negative logarithmic likelihood functions, they can also be used to perform Bayesian inference by combining with prior information on $h_1$ and $h_2$. We find that the results based on $S$ and $S_0$ are consistent, especially for big samples ($n \gtrsim 1000$). We suggest using $S$ for samples of $n < 1000$, and otherwise using $S_0$. Our considerations are as follows: (1) the influence of failing to observe a object with $L > \max(L_i)$ for small samples is larger than that for big samples, and setting a more conservative (small) $L_{\max}$ is necessary for small samples. Thus $S$ is more suitable for small samples as the double integration term provides extra constraint; (2) for big samples, dropping the double integration term can significantly save computational time and has little effect on the result. In addition, for the case of having difficulties in convergence while minimizing $S_0$, we suggest trying $S$. Finally, the KDE of the LF may be obtained by

$$\hat{\phi}(z, L) = \frac{n(Z_2 - Z_1)\hat{f}(x, y|h_1, h_2)}{(z - Z_1)(Z_2 - z)\Omega \frac{dV}{dz}}. \tag{15}$$

### 2.4. *Adaptive Kernel Density Estimation*

In Equation (9) the bandwidths are constant for every individual kernel. This may suffer from a slight drawback when applied to the typical $L-z$ data which exhibit inhomogeneous point patterns. The fixed bandwidths, even being chosen "optimally" in some sense, can not balance the estimates for sparsely and densely populated areas simultaneously. There is a tendency for spurious noise to appear for the former; if the estimates are smoothed sufficiently to deal with this, then oversmoothing would occur for the latter (e.g., Silverman 1986; Davies & Baddeley 2018). An effective solution to this problem is using variable bandwidths or adaptive kernel estimator, which allows the bandwidths of the kernels to vary from one point to another. The adaptive KDE have been shown to possess both theoretical and practical advantages over their fixed-bandwidth counterpart (Abramson 1982; Davies et al. 2018). Following Paper I, we implement the adaptive KDE in the following steps:

1. Make a pilot estimate using Equations (7) to (14), and obtain the optimal bandwidth, denoted as $\tilde{h}_1, \tilde{h}_2$.

2. Let the bandwidth vary with the local density:

$$\begin{cases} \lambda_1(x_j, y_j) = h_{10}\tilde{f}(x_j, y_j|\tilde{h}_1, \tilde{h}_2)^{-\beta} \\ \lambda_2(x_j, y_j) = h_{20}\tilde{f}(x_j, y_j|\tilde{h}_1, \tilde{h}_2)^{-\beta}, \end{cases} \tag{16}$$

where $\tilde{f}(x_j, y_j|\tilde{h}_1, \tilde{h}_2)$ is calculated via Equation (9) given $h_1 = \tilde{h}_1$ and $h_2 = \tilde{h}_2$; $h_{10}$ and $h_{20}$ are global bandwidths, and $\beta$ is the sensitivity parameter satisfying $0 \leqslant \beta \leqslant 1$ (Silverman 1986).

3. In Equations (9) and (13), we replace $h_1$ and $h_2$ with $\lambda_1(x_j, y_j)$ and $\lambda_2(x_j, y_j)$, respectively. We can obtain the adaptive KDE for the density of $(x, y)$, denoted as $\hat{f}_a(x, y|h_{10}, h_{20}, \beta)$, and its leave-more-out estimator, denoted as $\hat{f}_{a,(-i)}(x, y|h_{10}, h_{20}, \beta)$. The expression of $\hat{f}_a$ and $\hat{f}_{a,(-i)}$ are given in Appendix A.

4. The adaptive KDE for the density of $(z, L)$ and the corresponding leave-more-out estimator are

$$\begin{cases} \hat{p}_a(z, L|h_{10}, h_{20}, \beta) = \dfrac{\hat{f}_a(x, y)(Z_2 - Z_1)}{(z - Z_1)(Z_2 - z)}, \\ \hat{p}_{a,(-i)}(z_i, L_i|h_{10}, h_{20}, \beta) = \dfrac{\hat{f}_{a,(-i)}(x_i, y_i)(Z_2 - Z_1)}{(z - Z_1)(Z_2 - z)}. \end{cases} \tag{17}$$

5. In Equations (11) and (14), replacing $\hat{p}$ and $\hat{p}_{(-i)}$ with $\hat{p}_a$ and $\hat{p}_{a,(-i)}$, respectively, it is easy to obtain the $S$ or $S_0$ functions for the adaptive KDE. The optimal values of $h_{10}, h_{20}$ and $\beta$ can also be obtained.

6. Finally, the adaptive KDE of the LF is

$$\hat{\phi}_a(z, L) = \frac{n(Z_2 - Z_1)\hat{f}_a(x, y|h_{10}, h_{20}, \beta)}{(z - Z_1)(Z_2 - z)\Omega \frac{dV}{dz}}. \tag{18}$$

### 2.5. Small sample approximation

It is rather common in practice that the sample size is small ($n \lesssim 200$) in the interest redshift range $(Z_1, Z_2)$. In this situation the bivariate KDE method by Equation (15) can not play its advantages and may give a poor estimate. A possible solution is to approximate the two-dimensional problem to one-dimensional KDE problem. A transformation of the data is also required in the small sample case. But this is a little different from the one given in Equation (7), instead, we have

$$z = z, \text{ and, } l = L - f_{\lim}(z). \tag{19}$$

The domain of the new data set, $\{(z_1, l_1), (z_2, l_2), ...(z_n, l_n)\}$, is $\{Z_1 \leqslant z \leqslant Z_2, l > 0\}$. The density of $(z, l)$, denoted as $\hat{f}(z, l)$ can be approximated by KDE. As the redshift interval $(Z_1, Z_2)$ is small and $\phi$ changes little over $\Delta z$, we have

$$\hat{f}(z, l) \approx \frac{\hat{f}_l(l)}{Z_2 - Z_1}, \tag{20}$$

where $\hat{f}_l$ is the marginal probability density function that can be estimated by

$$\hat{f}_l(l) = \frac{1}{nh} \sum_{j=1}^{n} \left[ K_1\left(\frac{l - l_j}{h}\right) + K_1\left(\frac{l + l_j}{h}\right) \right], \tag{21}$$

where $l_j = L_j - f_{\lim}(z_j)$ corresponding to the $j$th object, $h$ is the bandwidth, and $K_1$ is the univariate kernel function given by

$$K_1(u) = \frac{1}{\sqrt{2\pi}} \exp(-\frac{1}{2}u^2). \tag{22}$$

According to the theory of probability transformation,

$$\hat{p}(z, L) \equiv \hat{f}(z, L - f_{\lim}(z)) = \frac{1}{(Z_2 - Z_1)nh} \sum_{j=1}^{n} \left[ K_1\left(\frac{L - f_{\lim}(z) - l_j}{h}\right) + K_1\left(\frac{L - f_{\lim}(z) + l_j}{h}\right) \right]. \tag{23}$$

The corresponding leave-one-out estimator is

$$\hat{p}_{-i}(z_i, L_i) = \frac{2}{(Z_2 - Z_1)(2n - 1)h} \left[ \sum_{\substack{j=1 \\ j \neq i}}^{n} K_1\left(\frac{l_i - l_j}{h}\right) + \sum_{j=1}^{n} K_1\left(\frac{l_i + l_j}{h}\right) \right]. \tag{24}$$

Combining Equations (23) and (24) with Equation (11), we can obtain the optimal bandwidth $h$. Finally, the KDE to the LF at the bin $\Delta z$ is estimated as

$$\hat{\phi}_1(z = z_0, L) = \hat{p}(z = z_0, L|\hat{h})n(\Omega \frac{dV}{dz})^{-1}, \tag{25}$$

where $z_0 = (Z_1 + Z_2)/2$, the subscript 1 in $\hat{\phi}_1$ means one dimensional approximation. Similar with the six steps introduced in Section 2.4, we can give the adaptive version of $\hat{\phi}_1$, denoting as $\hat{\phi}_{1a}$ (see Appendix B for details).

## 2.6. *Inclusion of the survey selection function*

More often a quasar survey may have a complicated selection function, and its selection probability is a complicated function of $z$, $M$, and the spectral energy distribution of the object (e.g., Fan et al. 2001; Kim et al. 2020). In this case, our KDE method has to be slightly modified to include the contribution from this selection function. A basic idea is to consider a weight for each data point when using KDE. For the $i$th object with reshift of $z_i$ and absolute magnitude of $M_i$, its weight $w_i \equiv 1/\mathcal{P}(z_i, M_i)$, where $\mathcal{P}$ is the selection function at $(z_i, M_i)$. Intuitively, a object with selection function of 0.5 is "like" two observations at that location (Richards et al. 2006). To perform the KDE, one need to first transform the data by

$$x = \ln\left(\frac{z - Z_1}{Z_2 - z}\right), \text{ and, } y = M - f_{\lim}(z). \tag{26}$$

Then the KDE to the new data set is

$$\hat{f}_w(x, y) = \frac{2}{2N_{\text{eff}} h_1 h_2} \sum_{j=1}^{n} w_j \left( K(\frac{x-x_j}{h_1}, \frac{y-y_j}{h_2}) + K(\frac{x-x_j}{h_1}, \frac{y+y_j}{h_2}) \right), \tag{27}$$

where $N_{\text{eff}}$ is the effective sample size given by $N_{\text{eff}} = \sum_{i=1}^{n} w_i$. The corresponding leave-more-out estimator is

$$\hat{f}_{w,(-i)}(x_i, y_i) = \frac{2}{(2N_{\text{eff}} - \eta_i) h_1 h_2} \left( \sum_{j \in J_i} w_j K(\frac{x_i-x_j}{h_1}, \frac{y_i-y_j}{h_2}) + \sum_{j \in J_i'} w_j K(\frac{x_i-x_j}{h_1}, \frac{y_i+y_j}{h_2}) \right). \tag{28}$$

where $J_i = \{j : x_i \neq x_j \text{ and } y_i \neq y_j, \text{ for } j = 1, 2, ..., n\}$, and $J_i' = \{j : x_i \neq x_j, \text{ for } j = 1, 2, ..., n\}$, for i=1,2,...n; $\eta_i$ denotes the total number of terms that are excluded from the two summations. Finally, the weighted version KDE LF is given by

$$\hat{\phi}(z, M) = \frac{N_{\text{eff}}(Z_2 - Z_1)\hat{f}_w(x, y|h_1, h_2)}{(z - Z_1)(Z_2 - z)\Omega\frac{dV}{dz}}. \tag{29}$$

Following the steps of section 2.4, it is easy to obtain the adaptive versions for Equations (27) and (28), whose expressions are given in Appendix C.

## 3. SIMULATION RESULTS

Hereafter we denote the LF estimated by Equation (15) as $\hat{\phi}$, and the small sample approximation by Equation (25) as $\hat{\phi}_1$. Their adaptive versions are denoted as $\hat{\phi}_a$ and $\hat{\phi}_{1a}$, respectively. To test the effectiveness of each estimator, we apply them to the 200 simulated samples of Paper I. These samples are flux-limited, with their flux limits randomly drawn between $10^{-2.5}$ and $10^{-0.5}$ Jy, and they share the same input LF (the model A RLF of Yuan et al. 2017). Their sample sizes range from 2000 to 40,000. For more details about the simulation, one can see Section 4.2 of Paper I. As described in Section 2.2, the range of $(Z_1, Z_2)$ can be flexibly expanded or narrowed according to the practical need. To test the performance of our estimators for different $(Z_1, Z_2)$ ranges, we discuss two cases where samples are divided or not diveded into redshift bins.

### 3.1. *Dividing redshift bins*

We divide each simulated sample into eight redshift bins, i.e., $(Z_1, Z_2)$ takes $(0.0, 0.2)$, $(0.2, 0.5)$, $(0.5, 1.0)$, $(1.0, 1.7)$, $(1.7, 2.5)$, $(2.5, 3.5)$, $(3.5, 4.5)$, $(4.5, 6.0)$ in turn. We then apply our $\hat{\phi}$ and $\hat{\phi}_1$ estimators and also their adaptive versions ($\hat{\phi}_a$ and $\hat{\phi}_{1a}$) to each redshift bin for the 200 simulated samples. There are $8 \times 200 = 1600$ calculations for each LF estimator. For comparison, the result measured by the traditional binning method (denoted as $\hat{\phi}_{\text{bin}}$) of Page & Carrera (2000) is also given, where the same redshift binning is adopted. The choice of luminosity bin size is also important to $\hat{\phi}_{\text{bin}}$. Intuitively, neither too large or too small $L$ bins can give desired estimates. Choosing the right number of bins involves finding a good tradeoff between bias and variance (Wasserman 2006). There should exist an optimal $L$ bin size, which may mainly depends on the density distribution of $L$ but not simply on the sample size. But in practice, it is difficult to find the "optimal" $L$ bin size. We thus use an uniform $L$ bin size $\Delta L = 0.3$, also adopted by the many authors in literature (e.g., Page & Carrera 2000; Richards et al. 2006).

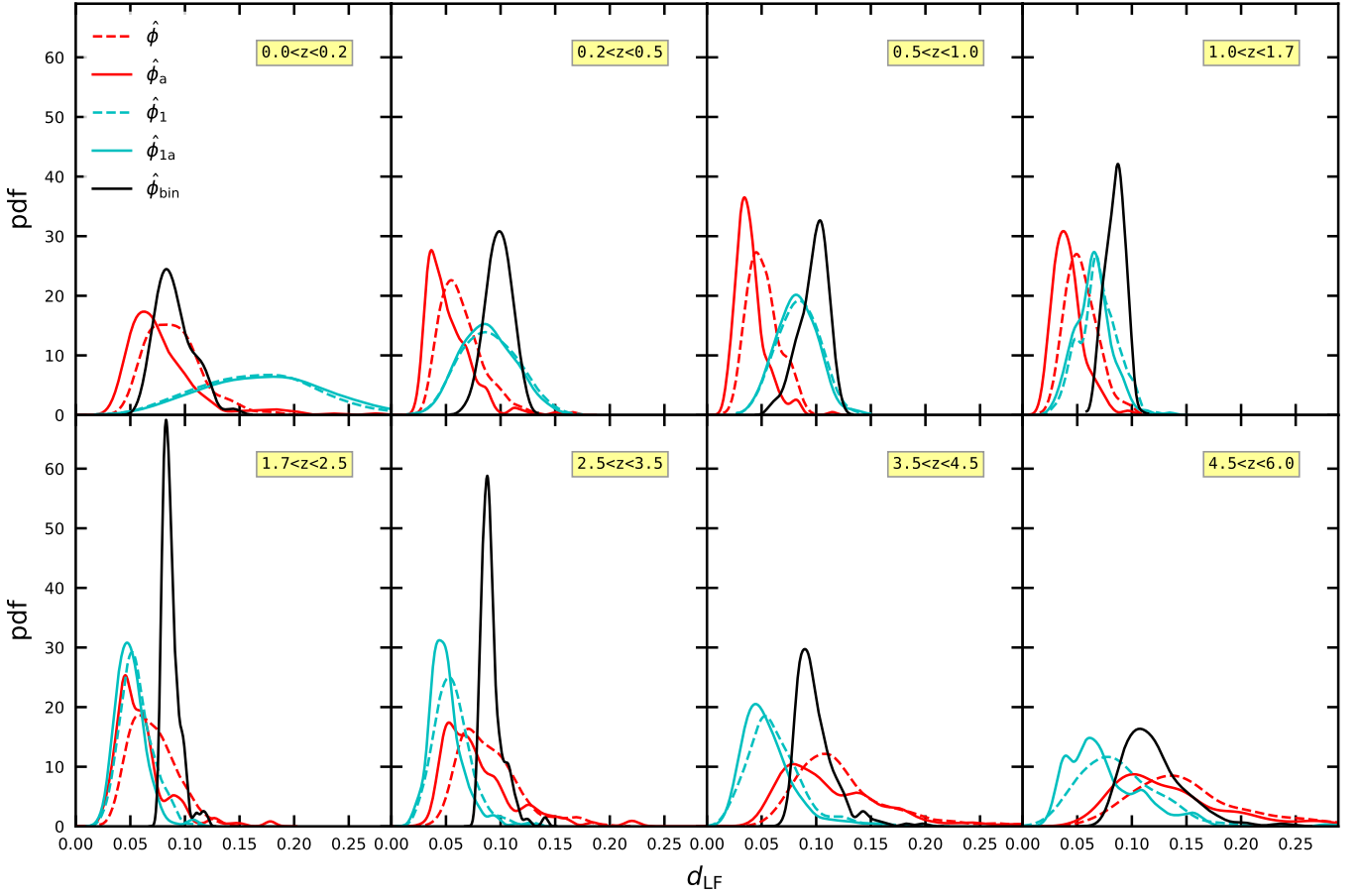

**Figure 3.** Distributions of $d_{\rm LF}$ in eight redshift bins for different LF estimators based on 200 simulated samples. In each redshift bin, the black solid curve represents the $\hat{\phi}_{\rm bin}$ estimator of Page & Carrera (2000). An uniform $L$ bin of $\Delta L = 0.3$ is used for $\hat{\phi}_{\rm bin}$. The red dashed and solid curves respectively correspond to the $\hat{\phi}$ and $\hat{\phi}_{\rm a}$ estimators; the cyan dashed and solid curves respectively represent the $\hat{\phi}_1$ and $\hat{\phi}_{\rm 1a}$ estimators.

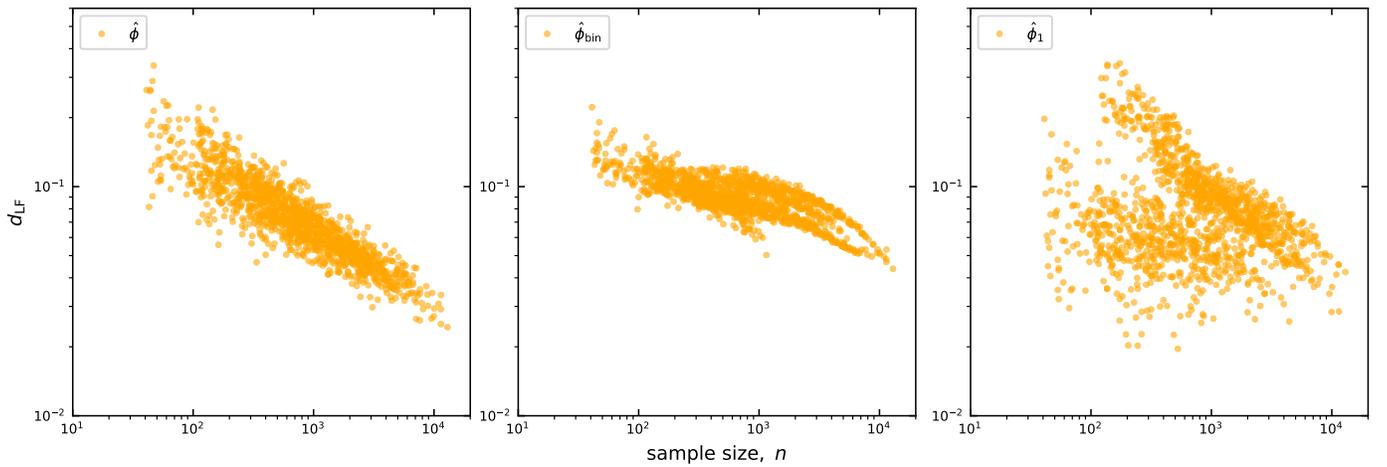

**Figure 4.** The statistic $d_{\rm LF}$ changes with the sample size $n$ for $\hat{\phi}$, $\hat{\phi}_{\rm bin}$ and $\hat{\phi}_1$ based on 200 simulated samples. Each sample is divided into eight redshift bins (edges of 0.0, 0.2, 0.5, 1.0, 1.7, 2.5, 3.5, 4.5, 6.0), thus there are 1600 $d_{\rm LF}$ value points in each panel. Variable $L$ bins of $\Delta L \propto 1/\log_{10}(n)$ are used for $\hat{\phi}_{\rm bin}$.

To quantify the performance of each LF estimator, following Paper I, we define a statistic $d_{\rm LF}$ for the $i$th redshift bin which measures the discrepancy of the estimated LF $\hat{\phi}$ from the true LF $\phi$ (it is known for the simulated samples). For the $\hat{\phi}$ (and $\hat{\phi}_{\rm a}$) estimator, the statistic is given by

$$d_{\rm LF}(\hat{\phi}) = \frac{1}{n_i} \sum_{j=1}^{n_i} \left| \log_{10} \left( \frac{\phi(z_j, L_j)}{\hat{\phi}(z_j, L_j)} \right) \right|, \qquad (30)$$

$n_i$ counts the number of objects in the $i$th redshift bin.

For the $\hat{\phi}_{\rm bin}$ estimator, estimates are given only at some discontinuous points, i.e., the centers of each $zL$ bin. The mathematics behind $\hat{\phi}_{\rm bin}$ is the histogram. In principle, a histogram can be considered as a step function that can be evaluated at an arbitrary point. This means that any point of in the $zL$ bin has the same LF, i.e., $\hat{\phi}_{\rm bin}(z_j, L_j) \equiv \hat{\phi}_{\rm bin}(z^c, L^c)$, where $(z^c, L^c)$ locates the center of the $zL$ bin. Therefore, Equation (30) should be applicable also for the $\hat{\phi}_{\rm bin}$ estimator.

For the $\hat{\phi}_1$ (and $\hat{\phi}_{\rm 1a}$) estimator, the statistic is given by

$$d_{\rm LF}(\hat{\phi}_1) = \frac{1}{n_i} \sum_{j=1}^{n_i} \left| \log_{10} \left( \frac{\phi(z=z_0, L_j)}{\hat{\phi}_1(z=z_0, L_j)} \right) \right|, \qquad (31)$$

where $z_0 = (Z_1 + Z_2)/2$. Note that the above calculation should exclude those few points with $L_j < f_{\rm lim}(z_0)$.

The statistic $d_{\rm LF}$ can be understood as the typical error of a LF estimator. So, the smaller the $d_{\rm LF}$ value the better. Figure 3 shows the distributions of $d_{\rm LF}$ in eight redshift bins for different LF estimators based on the 200 simulated samples. In each redshift bin, the black solid curve represents the $\hat{\phi}_{\rm bin}$ estimator. The red dashed and solid curves correspond to the $\hat{\phi}$ and $\hat{\phi}_{\rm a}$ estimators, respectively. In Table 1, we summarize the median $d_{\rm LF}$ for different LF estimators based on the 200 simulated samples. Except for the two highest redshift bins, the statistic $d_{\rm LF}$ of $\hat{\phi}$ and $\hat{\phi}_{\rm a}$ are better than that of $\hat{\phi}_{\rm bin}$. Moreover, the smaller dispersion of distributions suggests that they all have better stability than the $\hat{\phi}_{\rm bin}$ estimator, as expected. Comparing $\hat{\phi}$ and $\hat{\phi}_{\rm a}$, we find that generally the latter performs better, indicating that an adaptive KDE procedure is worthwhile.

Note that for the two highest redshift bins, our $\hat{\phi}$ and $\hat{\phi}_{\rm a}$ estimators do not exhibit advantages comparing with $\hat{\phi}_{\rm bin}$. This is because the sample size is relatively small ($\sim 50-300$) in the two highest redshift bins. For such a small sample size, a bivariate KDE method can not play its advantages. While this is a good opportunity for the $\hat{\phi}_{\rm 1a}$ estimator (shown as cyan solid curves in Figure 3) to play its value.

In Figure 4, we show how the statistic $d_{\rm LF}$ changes with the sample size $n$ for the $\hat{\phi}$, $\hat{\phi}_{\rm bin}$ and $\hat{\phi}_1$ estimators. In each panel, there are 1600 points which correspond to the $d_{\rm LF}$ values for the eight redshift bins of the 200 simulated samples. For our $\hat{\phi}$ estimator, there is significant inverse correlation between $d_{\rm LF}$ and $n$, indicating that it converges to the true LF quickly as $n$ increases. The inverse correlation for $\hat{\phi}_{\rm bin}$ is also obvious, but the trend is less steep. Thus we conclude that our $\hat{\phi}$ estimator converges to the true LF remarkably faster than the traditional $\hat{\phi}_{\rm bin}$ estimator. This conclusion is mathematically supported by Wasserman (2006). They proved that the kernel density estimators converge to the true density at the rate $O(n^{-4/5})$, faster than the histogram (rate of $O(n^{-2/3})$). As for the $\hat{\phi}_1$ estimator, there is not an obvious inverse correlation between $d_{\rm LF}$ and $n$. This is not surprising, since $\hat{\phi}_1$ makes some approximations, which inevitability leads to information loss. With the increase of $n$, the bias due to such information loss will dominate. In practice, $\hat{\phi}_1$ (and $\hat{\phi}_{\rm 1a}$) is mainly applied to the small $n$ cases ($\lesssim 200$), where it performs better than $\hat{\phi}$ (and $\hat{\phi}_{\rm a}$).

### 3.2. *No dividing redshift bins*

No dividing redshift bins corresponds to setting $(Z_1, Z_2)$ as $(0.0, 6.0)$ for the simulated samples. This means considering each sample as a whole and estimating a global LF in the redshift range of $(0.0, 6.0)$. For this "global LF" calculated by $\hat{\phi}$ (and $\hat{\phi}_{\rm a}$), we calculate its statistic $d_{\rm LF}$ respectively for the eight redshift bins defined in section 3.1. So we can judge the performance of the "global LF" by $\hat{\phi}$ (and $\hat{\phi}_{\rm a}$) at difference redshift intervals, and also compare with $\hat{\phi}_{\rm bin}$. To test the performance of $\hat{\phi}_{\rm bin}$ with different $L$ bins, we consider two cases: (1) adopting an uniform $L$ bin of $\Delta L = 0.3$, and (2) using variable $L$ bins of $\Delta L \propto 1/\log_{10}(n)$, where $n$ is the sample size in a redshift bin. Table 1 reports the median $d_{\rm LF}$ for $\hat{\phi}$ and $\hat{\phi}_{\rm a}$. They are significantly smaller than that for the cases of dividing redshift bins. Therefore, we emphasize that redshift binning is a matter of expediency for our $\hat{\phi}$ and $\hat{\phi}_{\rm a}$ estimators.

Note that when setting $(Z_1, Z_2)$ as $(0.0, 6.0)$, the estimate given by $\hat{\phi}_{\rm a}$ should be equivalent to the result of $\hat{\phi}_{\rm tra}$ in Paper I. In the Figures 7 and 8 of Paper I., the LFs estimated by $\hat{\phi}_{\rm tra}$ and $\hat{\phi}_{\rm bin}$, as well as the true LF, were shown.

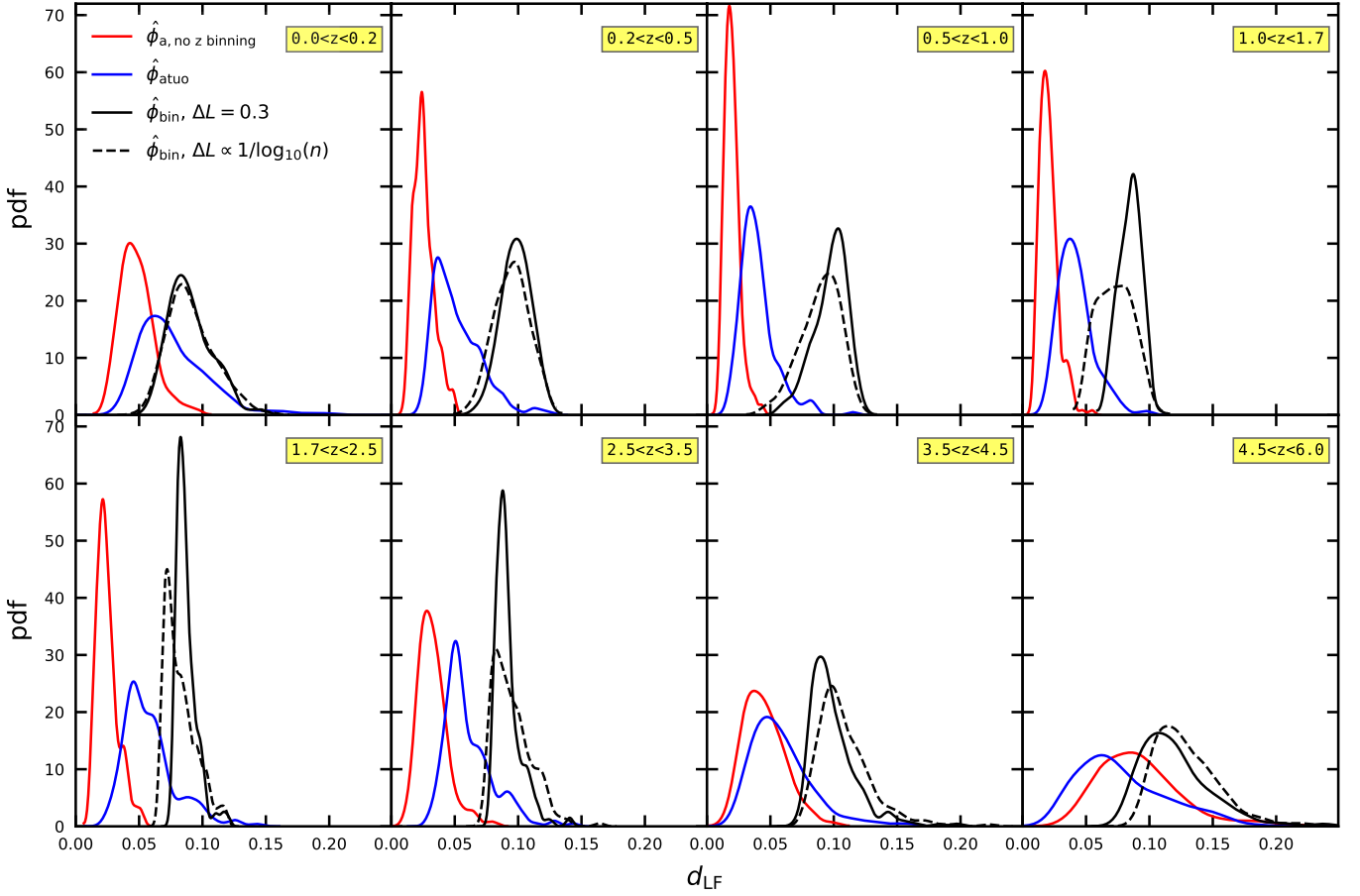

**Figure 5.** Distributions of $d_{\rm LF}$ in eight redshift bins for different LF estimators based on 200 simulated samples. The red solid curves represent $\hat{\phi}_{\rm a}$ with no dividing redshift bins. The black solid and dashed curves represent the results of uniform $L$ bin $\Delta L = 0.3$ and variable $L$ bins $\Delta L \propto 1/\log_{10}(n)$, respectively, for $\hat{\phi}_{\rm bin}$. The blue solid curves show the results of $\hat{\phi}_{\rm auto}$, which represents the optimal one chosen by the KS-test criterion among $\hat{\phi}$, $\hat{\phi}_{\rm a}$, $\hat{\phi}_1$ and $\hat{\phi}_{\rm 1a}$.

We refer the interested reader to there for an illustration. Figure 5 shows the distributions of $d_{\rm LF}$ for the $\hat{\phi}_{\rm a}$ and $\hat{\phi}_{\rm bin}$ estimators in eight redshift bins. $\hat{\phi}_{\rm a}$ performs significantly better than $\hat{\phi}_{\rm bin}$ in all the redshift bins. The two cases of uniform $L$ bin and variable $L$ bins for $\hat{\phi}_{\rm bin}$ are shown as black solid and dashed curves, respectively. It seems that adopting variable $L$ bins does not improve the performance of $\hat{\phi}_{\rm bin}$, indicating that the "optimal" $L$ bin size may mainly depends on the density distribution of $L$ but not simply on the sample size. Therefore, in practice it is very difficult to find some "optimal" $L$ bin size (and also redshift binning) for $\hat{\phi}_{\rm bin}$. The difficulties in $\hat{\phi}_{\rm bin}$ stems from the drawbacks of its mathematical foundation, i.e., the histogram. In the mathematical community, the histogram has been an outdated density estimator, used only for rapid visualization of results in one or two dimensions (e.g., Gramacki 2018).

## 4. DISCUSSIONS

For the case of dividing redshift bins, we have four LF estimators, $\hat{\phi}$, $\hat{\phi}_{\rm a}$, $\hat{\phi}_1$ and $\hat{\phi}_{\rm 1a}$. One may be confused about when and which estimator to use. Generally, the adaptive versions perform better, i.e., $\hat{\phi}_{\rm a}$ is better than $\hat{\phi}$, and $\hat{\phi}_{\rm 1a}$ is better than $\hat{\phi}_1$. At small sample size, $\hat{\phi}_1$ and $\hat{\phi}_{\rm 1a}$ generally performs better than $\hat{\phi}$ and $\hat{\phi}_{\rm a}$. However, there are some exceptions to the above qualitative criterion. Take $\hat{\phi}$ and $\hat{\phi}_{\rm a}$ for example, about 20% $d_{\rm LF}(\hat{\phi}_{\rm a})$ have values larger than that of $d_{\rm LF}(\hat{\phi})$ in our simulation (see Figure 6). Also note that there are some outliers which are poor estimates by $\hat{\phi}_{\rm a}$ and should be avoided. Therefore, we need a more quantitative criterion to decide which estimator to be used in a specific problem.

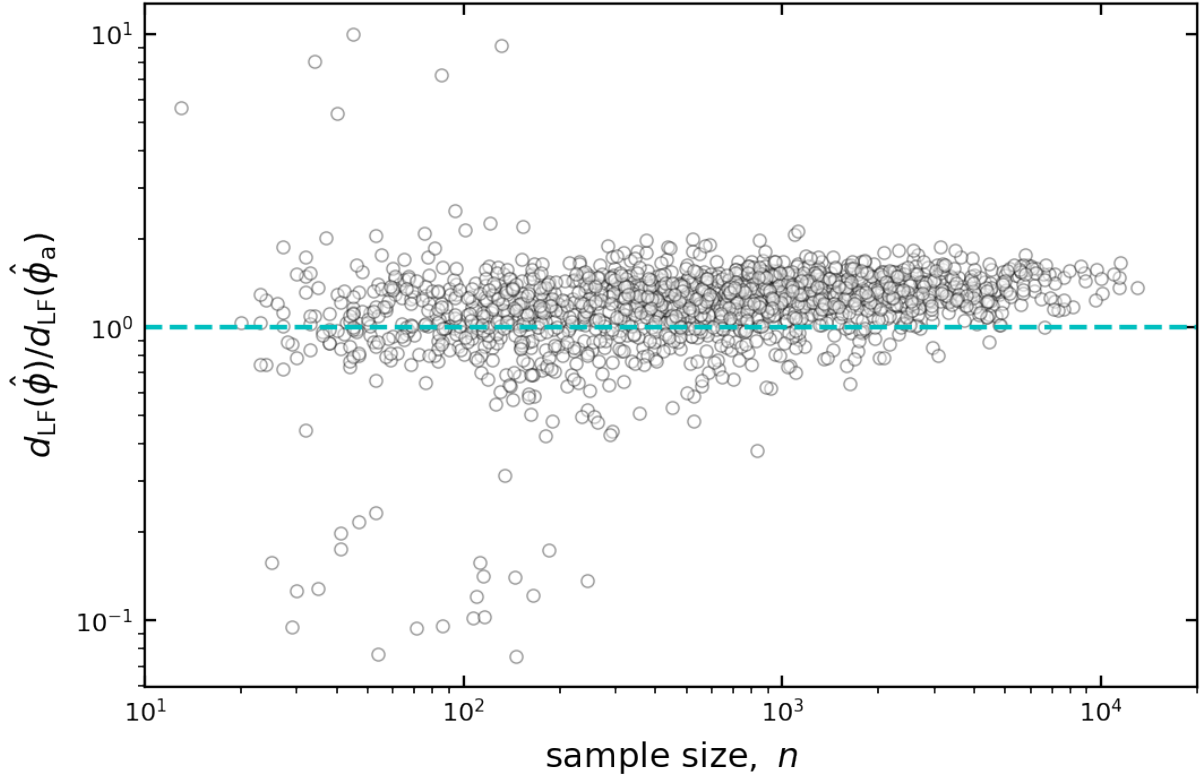

**Figure 6.** Ratios of $d_{\rm LF}(\hat{\phi})$ to $d_{\rm LF}(\hat{\phi}_{\rm a})$ as a function of sample sizes $n$ for the 200 simulated samples. The cyan dashed line indicates the boundary where the ratio equal to 1. About 20% points are below the boundary.

**Table 1.** Median $d_{\rm LF}$

| M($d_{\rm LF}$) \ z bin estimator | (0.0, 0.2) | (0.2, 0.5) | (0.5, 1.0) | (1.0, 1.7) | (1.7, 2.5) | (2.5, 3.5) | (3.5, 4.5) | (4.5, 6.0) | dividing redshift bins or not |
|---|---|---|---|---|---|---|---|---|---|
| $\hat{\phi}_{\rm bin}$ ($\Delta$ L=0.3) | 0.161 | 0.141 | 0.119 | 0.087 | 0.099 | 0.096 | 0.102 | 0.119 | |
| $\hat{\phi}$ | 0.088 | 0.060 | 0.050 | 0.052 | 0.069 | 0.085 | 0.114 | 0.147 | |
| $\hat{\phi}_{\rm a}$ | 0.071 | 0.047 | 0.037 | 0.040 | 0.055 | 0.072 | 0.107 | 0.134 | Yes |
| $\hat{\phi}_1$ | 0.169 | 0.089 | 0.083 | 0.069 | 0.054 | 0.055 | 0.058 | 0.084 | |
| $\hat{\phi}_{\rm 1a}$ | 0.179 | 0.087 | 0.082 | 0.064 | 0.049 | 0.048 | 0.051 | 0.068 | |
| $\hat{\phi}_{\rm auto}$ | 0.071 | 0.047 | 0.037 | 0.040 | 0.054 | 0.055 | 0.054 | 0.073 | |
| $\hat{\phi}$ | 0.073 | 0.035 | 0.032 | 0.034 | 0.036 | 0.052 | 0.064 | 0.149 | |
| $\hat{\phi}_{\rm a}$ | 0.047 | 0.024 | 0.019 | 0.019 | 0.023 | 0.031 | 0.043 | 0.088 | No |

**Notes**. Median $d_{\rm LF}$ (denoted as M($d_{\rm LF}$)) for different LF estimators based on the 200 simulated samples.

### 4.1. *Choosing the optimal estimator*

One appealing feature of KDE method is its continuous and smooth estimate. In Equations (10) and (15), once knowing the values for $h_1$ and $h_2$ by numerical approach, the function forms for $\hat{\phi}(z, L)$ and $\hat{p}(z, L)$ can be uniquely determined. Therefore the Kolmogorov-Smirnov test (KS-test) can measure the goodness-of-fit of $\hat{\phi}$ to the data. The classical one-dimensional KS-test compare the empirical and expected cumulative distribution functions (CDFs), $F_n(x)$ and $F(x)$. The KS-test statistic, $D_{\rm KS}$, is the maximum absolute difference between $F_n(x)$ and $F(x)$ for the same $x$:

$$D_{\rm KS} = \sup_{-\infty < x < \infty} |F_n(x) - F(x)|. \tag{32}$$

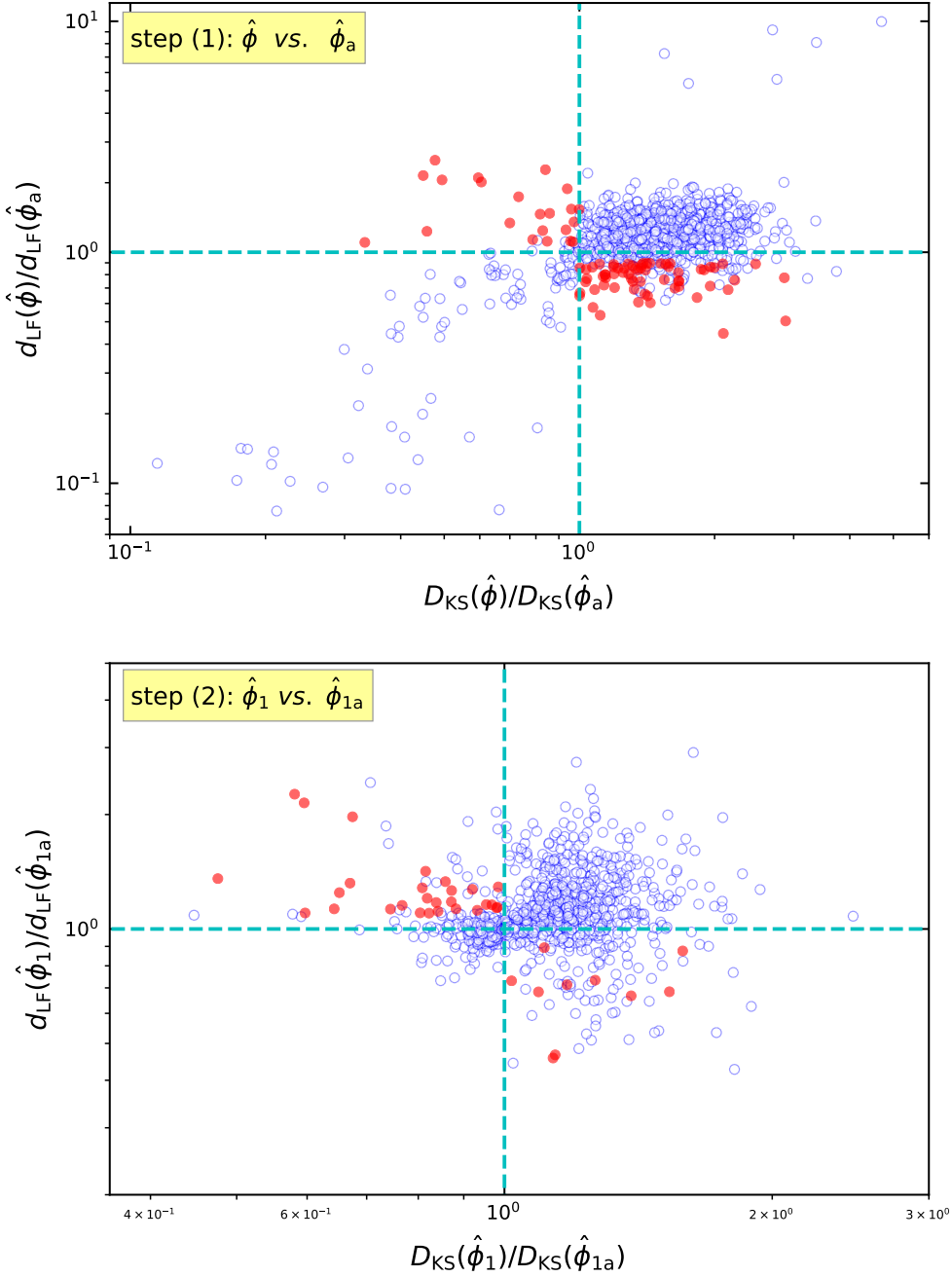

**Figure 7.** The relations between $d_{\rm LF}$ ratios and $D_{\rm KS}$ ratios for the first two steps, $\hat{\phi}$ vs. $\hat{\phi}_{\rm a}$, and $\hat{\phi}_1$ vs. $\hat{\phi}_{1\rm a}$. In each panel the cyan dashed lines indicate the boundary where the ratio equal to 1. The red solid circles mark the points that meet none of the ① to ④ conditions in section 4.1.

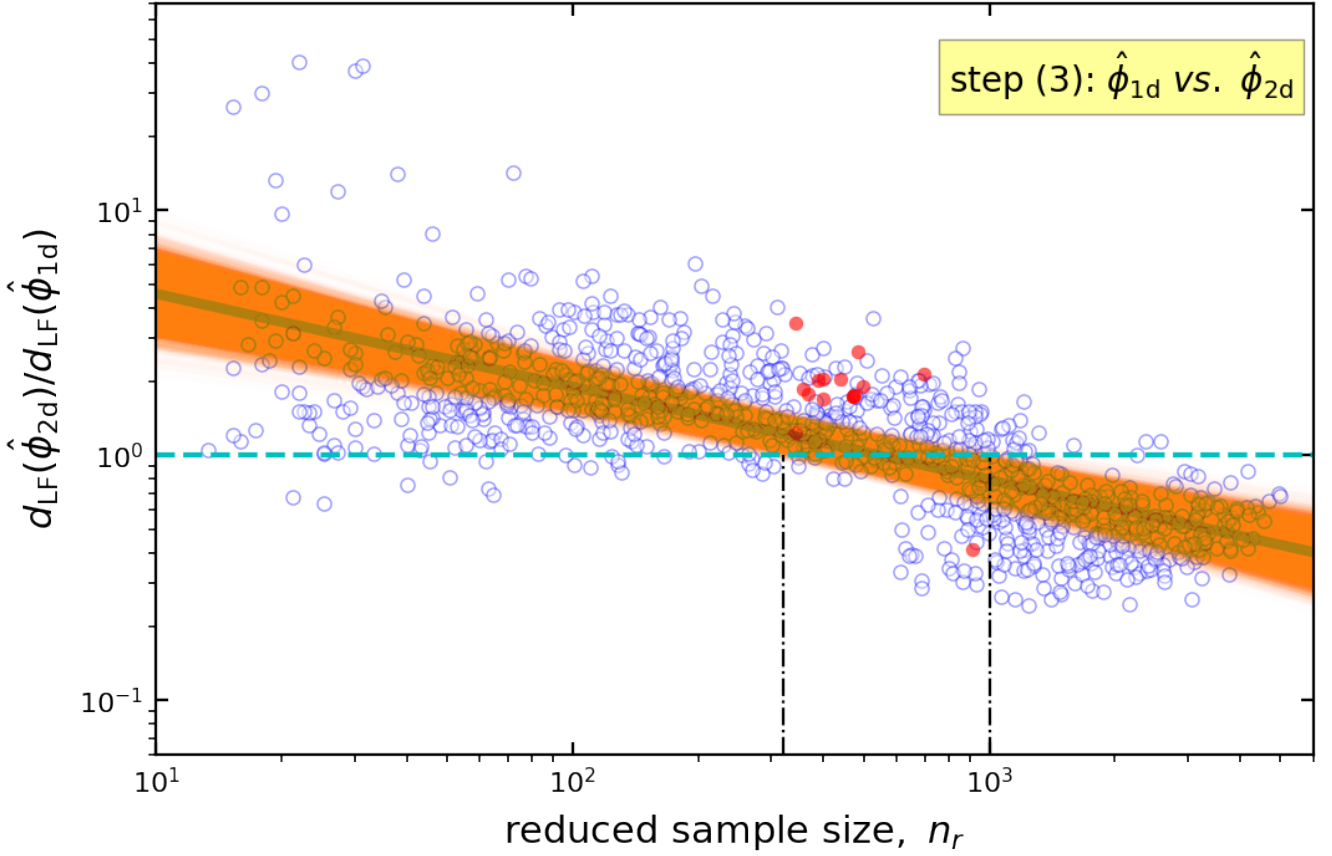

**Figure 8.** The $d_{\rm LF}$ ratios as a function of reduced sample sizes $n_r$ for the step (3), $\hat{\phi}_{\rm 1d}$ vs. $\hat{\phi}_{\rm 2d}$. The cyan dashed line indicates the boundary where the ratio equal to 1, and the red solid circles mark the points that meet none of the ① to ⑤ conditions in section 4.1. The black dash dot lines mark the critical $n_r$ of 320 and 1000. The gray thick solid line is the power-law fitting of $10.96 n_r^{-0.38}$, and the orange shaded area takes into account the $3\sigma$ error band.

For the $\hat{\phi}$ estimator, the expected CDF is given by

$$F(L) = \int_{Z_1}^{Z_2} \int_{f_{\rm lim}(z)}^{L} \hat{p}(z,L|h_1,h_2) dz dL, \tag{33}$$

where $\hat{p}(z,L)$ is given by Equation (10). The expected CDF for $\hat{\phi}_{\rm a}$, $\hat{\phi}_1$ and $\hat{\phi}_{\rm 1a}$ can be calculated similarly, just replacing $\hat{p}(z,L)$ with their own $\hat{p}$.

When one is confused about how to choose among $\hat{\phi}$, $\hat{\phi}_{\rm a}$, $\hat{\phi}_1$ and $\hat{\phi}_{\rm 1a}$, the KS-test will provide a criterion. One needs to first calculate LFs using the four estimators, and obtain their own $D_{\rm KS}$ values by Equation (32), respectively. Then choose the estimator with the smallest $D_{\rm KS}$ value as the optimal one. Specifically, there are three steps: (1) compare $\hat{\phi}$ and $\hat{\phi}_{\rm a}$, recording the optimal one as [2] $\hat{\phi}_{\rm 2d}$; (2) compare $\hat{\phi}_1$ and $\hat{\phi}_{\rm 1a}$, recording the optimal one as $\hat{\phi}_{\rm 1d}$; (3) compare $\hat{\phi}_{\rm 1d}$ and $\hat{\phi}_{\rm 2d}$, recording the optimal one as $\hat{\phi}_{\rm auto}$;

Note that the KS-test criterion does not require a knowledge of the true LF. In order to prove its effectiveness, we apply it to the 200 simulated samples and check the consistency between $d_{\rm LF}$ and $D_{\rm KS}$ statistically. A good consistency can guarantee that the KS-test criterion is effective. For the eight redshift bins of 200 simulated samples, we make 4ESTIMATING LUMINOSITY FUNCTIONS VIA KDE      151600 calculations and thus have 1600 pairs $(d_{\rm LF}, D_{\rm KS})$ for each LF estimator. We only consider those with $n \leqslant 1000$, leaving 1061 pairs $(d_{\rm LF}, D_{\rm KS})$.

---

[2] Note that $\hat{\phi}_{\rm 2d}$ is only a virtual name that may correspond to $\hat{\phi}$ or $\hat{\phi}_{\rm a}$, depending on which one has the smaller $D_{\rm KS}$ value. Similarly for $\hat{\phi}_{\rm 1d}$ and $\hat{\phi}_{\rm auto}$.

Take $\hat{\phi}$ and $\hat{\phi}_{\rm a}$ for example, the relation between $D_{\rm KS}(\hat{\phi})/D_{\rm KS}(\hat{\phi}_{\rm a})$ and $d_{\rm LF}(\hat{\phi})/d_{\rm LF}(\hat{\phi}_{\rm a})$ is plotted in the top panel of Figure 7. We find that most points (located at the upper right and lower left quadrants in the figure) meet one of the following conditions:

$$\begin{cases} ① : d_{\rm LF}(\hat{\phi}) > d_{\rm LF}(\hat{\phi}_{\rm a}) \text{ and } D_{\rm KS}(\hat{\phi}) > D_{\rm KS}(\hat{\phi}_{\rm a}) \\ ② : d_{\rm LF}(\hat{\phi}) < d_{\rm LF}(\hat{\phi}_{\rm a}) \text{ and } D_{\rm KS}(\hat{\phi}) < D_{\rm KS}(\hat{\phi}_{\rm a}) \end{cases}$$

This indicates a good consistency between $d_{\rm LF}$ and $D_{\rm KS}$. For the rest points, we find that quite a few of them meet one of the following conditions:

$$\begin{cases} ③ : 0.9 < d_{\rm LF}(\hat{\phi})/d_{\rm LF}(\hat{\phi}_{\rm a}) < 1.1 \\ ④ : d_{\rm LF}(\hat{\phi}) < 0.1 \text{ and } d_{\rm LF}(\hat{\phi}_{\rm a}) < 0.1 \end{cases}$$

The condition ③ implies that $d_{\rm LF}(\hat{\phi})$ and $d_{\rm LF}(\hat{\phi}_{\rm a})$ are very close, while the condition ④ indicates that both $d_{\rm LF}(\hat{\phi})$ and $d_{\rm LF}(\hat{\phi}_{\rm a})$ are relatively small. Thus for these "neutral" data points, choosing either $\hat{\phi}$ or $\hat{\phi}_{\rm a}$ is acceptable. In the top panel of Figure 7, the points that meet none of the ① to ④ conditions are shown as red solid circles, and the miss rate of the KS-test criterion is approximately equal to the percentage of these points. The miss rate of the KS-test criterion for the other two steps, $\hat{\phi}_1$ vs. $\hat{\phi}_{1a}$ and $\hat{\phi}_{1d}$ vs. $\hat{\phi}_{2d}$, can also be estimated similarly. The only difference is that for $\hat{\phi}_{1d}$ vs. $\hat{\phi}_{2d}$, the last condition should also be involved:

$$⑤ : n_r < 320 \text{ and } n_r > 1000,$$

where $n_r$ is the reduced sample size, defined by

$$n_r \equiv \frac{n}{Z_2 - Z_1}. \tag{34}$$

Because for small reduced sample size of $n_r < 320$, we always choose $\hat{\phi}_{1d}$ as the optimal one, while for $n_r > 1000$ we always choose $\hat{\phi}_{2d}$. The reason will be explained latter. Table 2 summarizes the result of miss rate. Basically, the miss rate for all the three steps are acceptable. Figure 8 plots the $d_{\rm LF}$ ratio as a function of reduced sample size $n_r$ for step (3). Intriguingly, there is a obvious trend that the ratio of $d_{\rm LF}(\hat{\phi}_{2d})$ to $d_{\rm LF}(\hat{\phi}_{1d})$ decreases with the increase of $n_r$. A power-law fitting gives $d_{\rm LF}(\hat{\phi}_{2d})/d_{\rm LF}(\hat{\phi}_{1d}) = 10.96 n_r^{-0.38}$ (shown as the gray thick solid line). The orange shaded area takes into account the $3\sigma$ error band. The two outermost intersection points between the cyan dashed line and the shaded area are at $n_r \approx 320$ and $n_r \approx 1000$ (marked by the black dash dot lines). Below $n_r \approx 320$ nearly all the points have ratios of $d_{\rm LF}(\hat{\phi}_{2d})/d_{\rm LF}(\hat{\phi}_{1d}) > 1$, while above $n_r \approx 1000$ nearly all ratios are smaller than 1. This is where the condition ⑤ comes from. In summary, we have proved that the KS-test criterion is effective for choosing an optimal estimator among $\hat{\phi}$, $\hat{\phi}_{\rm a}$, $\hat{\phi}_1$ and $\hat{\phi}_{1a}$ in practice.

### 4.2. *A guidance for using our LF estimators*

We summarize a guidance for using our LF estimators as follows:

1. Do not divide redshift bins, only if the sample size is too large ($n > 10^5$).

2. If binning is essential, one should make every redshift bin contain sources as many as possible.

3. For small reduced sample size of $n_r \lesssim 320$, use the small sample approximation. The choice of $\hat{\phi}_1$ or $\hat{\phi}_{1a}$ depends on which one has the smaller $D_{\rm KS}$ value.

4. For small-medium reduced sample size of $320 < n_r < 1000$, use the KS-test criterion to choose the optimal one from $\hat{\phi}$, $\hat{\phi}_{\rm a}$, $\hat{\phi}_1$ and $\hat{\phi}_{1a}$.

5. For large reduced sample size of $n_r > 1000$, use $\hat{\phi}$ or $\hat{\phi}_{\rm a}$, depending on which one has the smaller $D_{\rm KS}$ value.

It should be emphasized that the critical values $n_r = 320$ and $n = 1000$ are experiential, and appropriate changes are allowed. We denote the "optimal LFs" obtained by the guidance 3 to 5 as $\hat{\phi}_{\rm auto}$. Figure 5 shows the distributions of $d_{\rm LF}$ in eight redshift bins for $\hat{\phi}_{\rm auto}$ based on the 200 simulated samples. In Table 1, we report the median $d_{\rm LF}$ for $\hat{\phi}_{\rm auto}$. On the whole, $\hat{\phi}_{\rm auto}$ combines the advantages of $\hat{\phi}$, $\hat{\phi}_{\rm a}$, $\hat{\phi}_1$ and $\hat{\phi}_{1a}$, and performs significantly better than $\hat{\phi}_{\rm bin}$ in all the eight redshift bins.

**Table 2.** Miss rate

|  | step (1) $\hat{\phi}$ vs. $\hat{\phi}_a$ | step (2) $\hat{\phi}_1$ vs. $\hat{\phi}_{1a}$ | step (3) $\hat{\phi}_{1d}$ vs. $\hat{\phi}_{2d}$ |
|---|---|---|---|
| miss rate | 8.20% | 3.58% | 1.41% |

**Notes**. Miss rate for the three steps of using the KS-test criterion, estimated based on the 200 simulated samples.

```
1 from kdeLF import kdeLF
2 test=kdeLF.KdeLF(sample_file=*,zbin=*,f_lim=*,...)
3 test.get_optimal_h()
4 test.get_lgLF()
5 test.KStest()
6 test.run_mcmc()
7 test.chain_analysis()
```

**Figure 9.** Importing `kdeLF` and its subroutines to illustrate the overall structure of `kdeLF`.

### 4.3. *Comparison with Paper I*

The KDE method in Paper I is only applicable to problems where the sample has a coverage of redshift as broad as possible, i.e., $Z_1 \simeq 0$ and $Z_2 \gg Z_1$. The KDE method in this paper is not subject to this restriction. In order to reduce boundary effects of KDE, Paper I introduced two solutions, the transformation and transformation–reflection methods. This work inherits the latter method, and add a consideration for the case of small sample size. In a word, this work has generalized our previous KDE method. This new upgrade greatly extend the application scope of our KDE method, making it a very flexible approach which is suitable to deal with nearly all kinds of bivariate LF estimating problems.

## 5. A PYTHON IMPLEMENTATION OF THE KDE METHOD

We have developed a software toolbox to implement our KDE method, called `kdeLF`. The `kdeLF` package is mainly written in Python and combines the Fortran language to resolve speed bottlenecks. `kdeLF`'s interface is designed to be simple, intuitive, and easily extensible.

### 5.1. *Package Structure and Basic Usage*

The overall structure of the `kdeLF` package is illustrated in Figure 9, where the package and its main subroutines are imported or executed in a typical python session. These subroutines have fairly obvious names: `get_optimal_h` obtains the optimal values for bandwidths using maximum likelihood estimation (MLE); `get_lgLF` gets the logarithm values of the LFs estimated via the KDE estimators, and also plots the LFs; `KStest` performs the KS-test and plots the empirical CDF and expected CDFs by the KDE estimators; `run_mcmc` implements a fully Bayesian Markov Chain Monte Carlo (MCMC) method to determine the posterior distributions of bandwidths and other parameters (e.g., $\beta$ for adaptive estimators). The MCMC core embedded in `kdeLF` is the Python package `emcee` (Foreman-Mackey et al. 2013). The Bayesian method allows us to recover the parameters with a complete description of their uncertainties and degeneracies via calculating their probability density functions (PDFs). Then, by running the `chain_analysis` subroutine, one can probe the shape of these PDFs, and the correlations among bandwidth parameters, giving more information than just the best-fit and the marginalized values for the parameters (also see Calistro Rivera et al. 2016).

`kdeLF.KdeLF` acts as the interface which receives the input of required and optional arguments to initialize a calculation. `KdeLF` instances can be initialized, e.g., as

```
from kdeLF import kdeLF
test=kdeLF.KdeLF(sample_file='data.txt',zbin=[0.0,0.2],
    f_lim=f_lim,solid_angle=0.5,H0=71,Om0=0.27,
    small_sample_approximation=False,adaptive=False) ,
```

```
1  import numpy as np
2  from kdeLF import kdeLF
3  from scipy import interpolate
4
5  with open('flim.dat', 'r') as f:
6      x0, y0= np.loadtxt(f, usecols=(0,1), unpack=True)
7  f_lim = interpolate.interp1d(x0, y0,fill_value="extrapolate")
8  sr=6248/((180/np.pi)**2)
9
10 lf=kdeLF.KdeLF(sample_name='data.txt',solid_angle=sr,zbin=[0.6,0.8],f_lim=f_lim,adaptive=True)
11 lf.get_optimal_h()
12 lf.run_mcmc()
13 lf.plot_posterior_LF(z=0.718,sigma=3)
```

**Figure 10.** Example of code to run `kdeLF` for the $0.6 < z < 0.8$ data.

which sets up an `KdeLF` instance, named `test`, that represents the initial conditions of a LF calculation. The first four arguments are required and the others are optional. `sample_file` is the name of sample file that contains at least two columns of data for $z$ and $L$ (or absolute magnitude $M$). If the user wish to calculate a KDE LF considering weighing, a third column containing selection probabilities should be provided in `sample_file`. `zbin` is the redshift range $[Z_1, Z_2]$. `f_lim` is the user defined Python function calculating the truncation boundary $f_{\lim}(z)$ of sample, and `solid_angle` is the solid angle (unit of $sr$) subtended by the sample. `kdeLF` adopts a Lambda Cold Dark Matter (LCDM) cosmology, but it is not limited to this specific cosmological model. The optional arguments `Om0` and `H0`, defaulting as 0.30 and 70 (km s$^{-1}$ Mpc$^{-1}$), represent the $\Omega_m$ parameter and Hubble constant for the LCDM cosmology, respectively. The optional arguments `small_sample_approximation` and `adaptive` have four different combinations, (`False`, `False`), (`False`, `True`), (`True`, `False`) and (`True`, `True`), corresponding to the usages of $\hat{\phi}$, $\hat{\phi}_a$, $\hat{\phi}_1$ and $\hat{\phi}_{1a}$, respectively.

The speed of the `kdeLF` code mainly depends on the sample size $n$. The most time-consuming part of the calculation in `kdeLF` is implemented by Fortran, and a parallel processing via OpenMP (Open Multi-Processing) is used. For a calculation of $n = 5000$, `kdeLF` takes about one minute to finish the `get_optimal_h` subroutine on a 8-core personal computer. `run_mcmc` is the most time-consuming subroutine, which takes about 1-2 hours.

### 5.2. An Example

As a test example, we apply `kdeLF` to the quasar sample compiled by Kulkarni et al. (2019). The sample is from the SDSS DR7 quasar catalogue (Schneider et al.2010). Kulkarni et al. (2019) restricted it to the 48 664 $0.1 < z < 2.2$ quasars selected with the final SDSS quasar selection algorithm (Richards et al. 2006) from a survey area of 6248 deg$^2$. Considering that the lower redshift bins are excluded from Kulkarni et al. (2019)'s LF evolution analysis due to obvious systematic errors, we only use the $0.6 < z < 2.2$ quasars. The sample is not assumed to be complete within its survey region. The $i$th quasar $(z_i, M_i)$ is associated with a value for the selection function $\mathcal{P}(z_i, L_i)$, which can be interpreted as the probability that a quasar is at this location. To use `kdeLF`, we prepare a txt file named, say 'data.txt'. The first four lines of its content

```
# z          M         P(z,M)
0.601      -23.29       0.9453
0.601      -22.32       0.9405
0.601      -22.42       0.9405
...
```

With this three-column data, `kdeLF` will automatically implement the KDE analysis considering the weighing due to the selection function. For comparison, we divided the sample into the same seven redshift bins as Kulkarni et al. (2019). For the data of each redshift bin, we can respectively run `kdeLF` through a few lines of simple code to obtain a KDE LF analysis. Figure 10 shows a example of code for the $0.6 < z < 0.8$ bin.

Figure 11 shows the resultant LF estimates by $\hat{\phi}_a$ for the seven redshift bins. These are shown as blue solid curves with orange error bands of $3\sigma$ (99.93 %) uncertainties. Different from the $\hat{\phi}_{\rm bin}$ estimator, our KDE method is able to give the LF values at any redshift in the range of $Z_1 < z < Z_2$. The LFs are plotted at the mean redshift of each redshift bin. The red solid circles indicate the binned LFs inferred by Kulkarni et al. (2019). Our KDE LFs are in good agreement with their estimates. But for the bright end LFs, our KDE estimates have smaller uncertainties. The main difference occurs at the faint limit, where the binned LFs show a suspicious decline (red open circles in the Figure). Kulkarni et al. (2019) thus concluded that the SDSS selection function is systematically overestimated at its magnitude limit. They treated those few bins near the faint limt as "discrepant" ones, and discarded the contributing quasars from their analysis. We argued that although the possibility of overestimation for the selection function can not be ruled out, the $\hat{\phi}_{\rm bin}$ method they used is also an important factor. Through a detailed graphical analysis and Monte Carlo simulation, Yuan & Wang (2013) concluded that once the LF is evolving with redshift, the classical binned methods will inevitability underestimate the density for the bins which are crossed by the flux limit curves. This is a typical boundary bias. In our KDE analysis, we do not discarded the "discrepant" quasars. Figure 11 shows that our KDE LFs do not show a suspicious decline at the faint limit. This is owe to the reflection operation in our KDE method, which can significantly minimize the boundary bias near the faint limit.

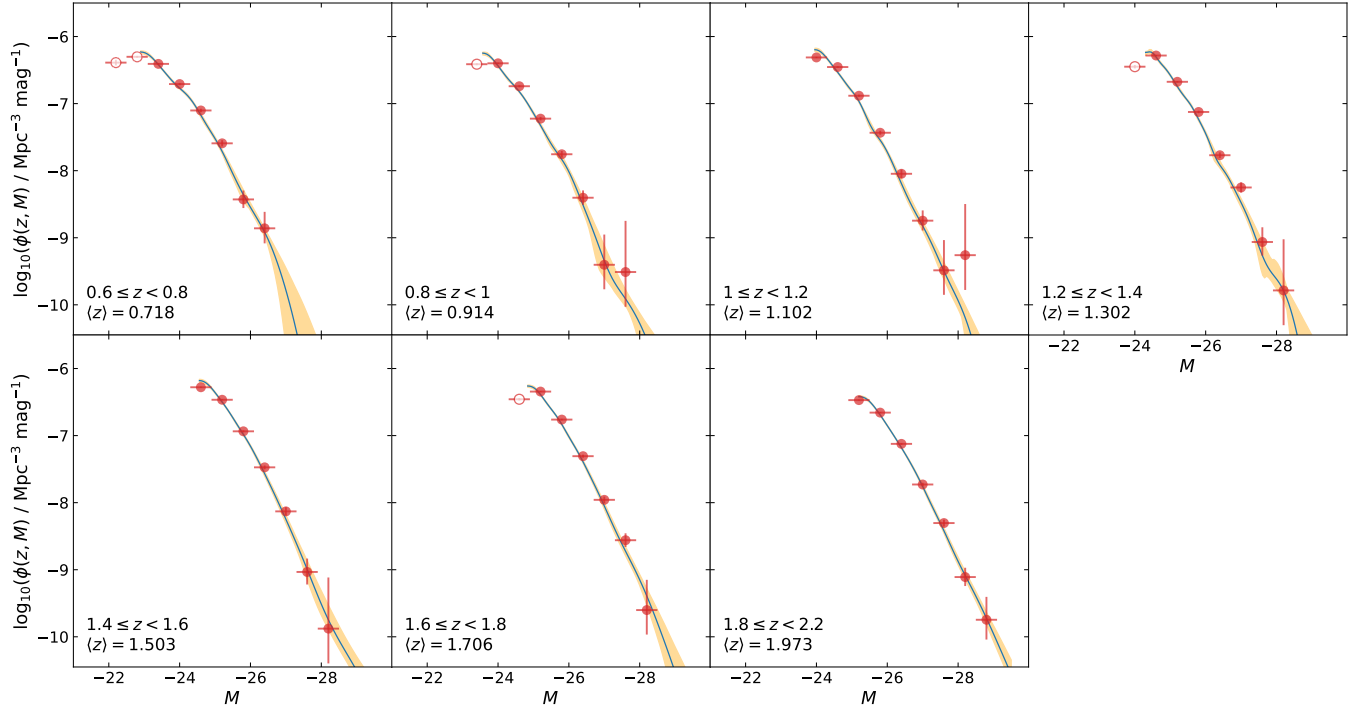

**Figure 11.** LF estimates from $z = 0.6$ to $2.2$. The blue curves show our fiducial LF given by $\hat{\phi}_a$. The shaded regions show the $3\sigma$ (99.93 %) uncertainties. The red solid circles indicate the binned LFs inferred by Kulkarni et al. (2019). The red open circles indicate magnitude bins excluded in their paper due to incompleteness in the data set.

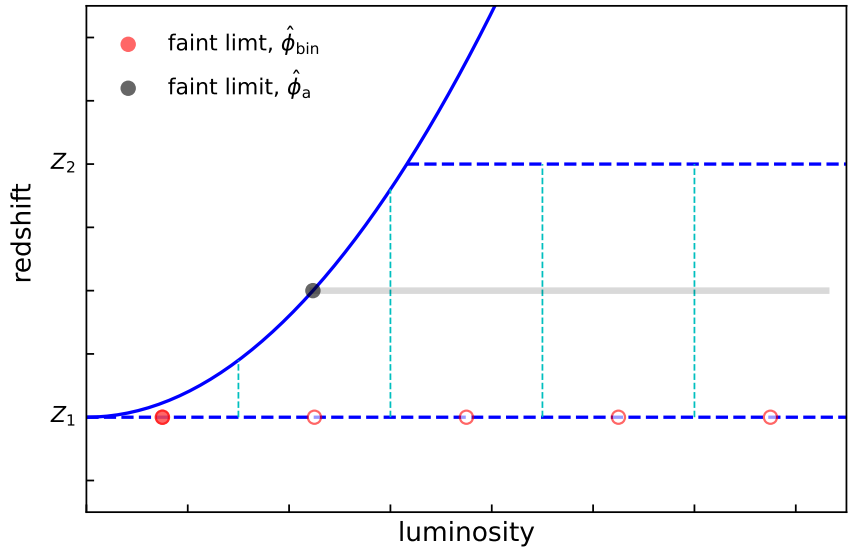

**Figure 12.** A sketch illustrating the faint luminosity limits of the LFs estimated by $\hat{\phi}_a$ and $\hat{\phi}_{\rm bin}$.

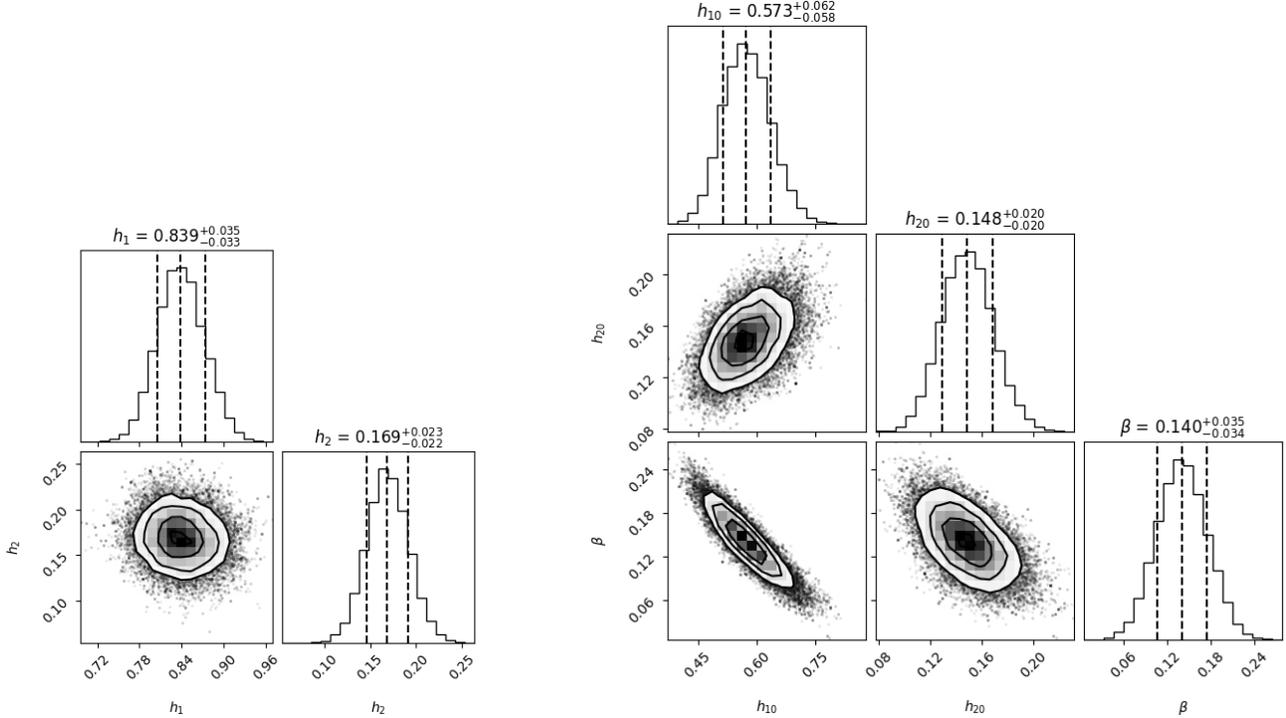

**Figure 13.** Posterior distributions of the bandwidth parameters for $\hat{\phi}$ (left) and $\hat{\phi}_a$ (right) in the $0.6 < z < 0.8$ bin. The vertical lines mark the locations of best-fit values and $1\sigma$ errors by MCMC.

A careful inspection may find that our KDE LFs do not extend down to the faint luminosities of the open circles in Figure 11. This is due to the existence of survey limit, the faint luminosity limits are different for different redshifts. Figure 12 illustrates how does this happen. $\hat{\phi}_{\rm bin}$ gives discrete LF points, whose faint luminosity limit is shown as red solid circle in Figure 12. $\hat{\phi}_a$ (and our other KDE estimators) gives continuous LF, whose faint luminosity limit is shown as black solid circle in the figure. If the LF by $\hat{\phi}_a$ takes a $z$ value close to $Z_1$, it can extend down to the faint luminosity of red solid circle.

One important feature of our KDE method is involving the Bayesian inference. Figure 13 shows marginalized 1D and 2D posterior probability distribution functions of the bandwidth parameters $h_1$ and $h_2$ for $\hat{\phi}$, and $h_{10}$, $h_{20}$ and $\beta$ for $\hat{\phi}_a$. The values of $h_1$ and $h_2$ are used as pilot bandwidths for implementing $\hat{\phi}_a$ (see Section 2.4). The Figure illustrates that all the five parameters are well constrained.

### 5.3. *Future Development*

In mathematics, KDE is a very successful smoothing technique with many practical applications. Some of its aspects, such as bandwidth selection, and fast KDE algorithm for big data, are still at the forefront of research. We are optimistic about the application prospect of KDE method in LF measurement and also other fields of astronomy. We anticipate that our method and the `kdeLF` code, especially the latter, will be under continuous development over the coming years, and that this upgrade will be driven by the needs of its users. The source code is currently hosted in a git repository on GitHub at http://github.com/yuanzunli/kdeLF, and version 1.0.0 is archived in Zenodo (Yuan et al. 2022). Users are encouraged to report bugs through the issue tracker on GitHub, and contributions are welcome. To learn more about how to use `kdeLF` in practice, it is best to check out the documentation on the website http://kdelf.readthedocs.org/en/latest , where the API documentation and some examples are presented.

## 6. SUMMARY

We summarize the important points of this work as follows.

1. We propose a generalization of our previous KDE (kernel density estimation) method for estimating luminosity functions (LFs). This new upgrade further extend the application scope of our KDE method, making it a very flexible approach which is suitable to deal with nearly all kinds of bivariate LF estimating problems.

2. We believe that, from the mathematical point of view, the LF calculation can be abstracted as a density estimation problem in the bounded domain of $\{Z_1 < z < Z_2, L > f_{\lim}(z)\}$. We use the transformation-reflection KDE method to solve the problem and denote the estimator as $\hat{\phi}$. For small sample size situation, an approximate method based on one-dimensional KDE is proposed, denoting as $\hat{\phi}_1$. To further improve their performance, we introduce their adaptive versions, denoting as $\hat{\phi}_a$ and $\hat{\phi}_{1a}$.

3. Based on 200 simulated samples, we find that for the cases of no dividing redshift bins, $\hat{\phi}$ and $\hat{\phi}_a$, especially the latter, have a very good performance, and have overwhelming superiority to the traditional binning method ($\hat{\phi}_{\rm bin}$).

4. For the cases of dividing redshift bins, we propose a KS-test criterion to choose the optimal LF estimator from $\hat{\phi}$, $\hat{\phi}_a$, $\hat{\phi}_1$ and $\hat{\phi}_{1a}$ according to a simple and easy-to-operate guidance. The selected estimator, denoted as $\hat{\phi}_{\rm auto}$, combines the advantages of $\hat{\phi}$, $\hat{\phi}_a$, $\hat{\phi}_1$ and $\hat{\phi}_{1a}$, and performs significantly better than $\hat{\phi}_{\rm bin}$ in all the eight redshift bins.

5. The simulation suggests that with the increase of $n$, our KDE LF estimator converges to the true LF remarkably faster than $\hat{\phi}_{\rm bin}$.

6. We have developed a public, open-source Python Toolkit, called `kdeLF`, to implement our KDE method. The performance of the code for real data was tested on a quasar sample from the SDSS. With the support of `kdeLF`, our KDE method could be a competitive alternative to existing nonparametric estimators, due to its high accuracy and excellent stability.

## ACKNOWLEDGMENTS

We thank the anonymous reviewer for the many valuable comments and suggestions, which significantly improve the presentation of paper. We acknowledge the financial support from the National Natural Science Foundation of China (grant No. 12073069), Yunnan Natural Science Foundation (Nos. 2019FB008 and 2019FB009), and the science research grants from the China Manned Space Project with NO.CMS-CSST-2021-A12, CMS-CSST-2021-B10. Z.Y. would like to thank Xian Hou and Guobao Zhang for providing computing platforms which promoted the progress of this work. `kdeLF` makes use of the open-source Python numpy package.

*Software:* SciPy (Virtanen et al. 2020), Astropy (Astropy Collaboration et al. 2013), emcee (Foreman-Mackey et al. 2013), corner (Handley 2018), GetDist (Lewis 2019), QUADPACK (Piessens et al. 1983) , zeus (Karamanis et al. 2021), matplotlib (Hunter 2007).

## APPENDIX

### A. THE ADAPTIVE KDE

The adaptive KDE is

$$\hat{f}_a(x,y) = \frac{1}{n}\sum_{j=1}^{n}\frac{1}{\lambda_1(x_j,y_j)\lambda_2(x_j,y_j)}\left\{K\left(\frac{x-x_j}{\lambda_1(x_j,y_j)},\frac{y-y_j}{\lambda_2(x_j,y_j)}\right)+K\left(\frac{x-x_j}{\lambda_1(x_j,y_j)},\frac{y+y_j}{\lambda_2(x_j,y_j)}\right)\right\}, \tag{A1}$$

where $\lambda_1(x_j,y_j)$ and $\lambda_2(x_j,y_j)$ are calculated via Equation (16). The leave-more-out estimator is

$$\hat{f}_{a,(-i)}(x_i,y_i) = \frac{2}{(2n-\eta_i)}\left\{\sum_{j\in J_i}\frac{K\left(\frac{x_i-x_j}{\lambda_1(x_j,y_j)},\frac{y_i-y_j}{\lambda_2(x_j,y_j)}\right)}{\lambda_1(x_j,y_j)\lambda_2(x_j,y_j)}+\sum_{j\in J'_i}\frac{K\left(\frac{x_i-x_j}{\lambda_1(x_j,y_j)},\frac{y_i+y_j}{\lambda_2(x_j,y_j)}\right)}{\lambda_1(x_j,y_j)\lambda_2(x_j,y_j)}\right\}, \tag{A2}$$

where $J_i = \{j : x_i \neq x_j \text{ and } y_i \neq y_j, \text{for } j=1,2,...,n\}$, and $J'_i = \{j : x_i \neq x_j, \text{for } j=1,2,...,n\}$, for i=1,2,...n; $\eta_i$ denotes the total number of terms that are excluded from the two summations.

### B. THE ADAPTIVE VERSION OF $\hat{\phi}_1$

The adaptive version of $\hat{\phi}_1$ can be implemented in the following steps:

1. Make a pilot estimate using Equations (19) to (25), obtaining the optimal bandwidth, denoted as $\tilde{h}$.

2. Let the bandwidth vary with the local density:

$$H(l) = h_0 \tilde{f}_l(l|\tilde{h})^{-\beta}, \tag{B3}$$

where $\tilde{f}_l(l|\tilde{h})$ is calculated via Equation (21) given $h = \tilde{h}$, $h_0$ is the global bandwidth, and $\beta$ is a free parameter.

3. In Equations (23) and (24), replace $h$ with $H(l_j)$. Then we can obtain the adaptive KDE $\hat{p}_a(z,L)$ and its leave-one-out estimator $\hat{p}_{a,-i}(z_i,L_i)$:

$$\hat{p}_a(z,L) = \frac{1}{(Z_2-Z_1)n}\sum_{j=1}^{n}\frac{1}{H(l_j)}\left[K\left(\frac{L-f_{\lim}(z)-l_j}{H(l_j)}\right)+K\left(\frac{L-f_{\lim}(z)+l_j}{H(l_j)}\right)\right], \tag{B4}$$

and

$$\hat{p}_{a,-i}(z_i,L_i) = \frac{2}{(Z_2-Z_1)(2n-1)}\left[\sum_{\substack{j=1\\j\neq i}}^{n}\frac{1}{H(l_j)}K\left(\frac{l_i-l_j}{H(l_j)}\right)+\sum_{j=1}^{n}\frac{1}{H(l_j)}K\left(\frac{l_i+l_j}{H(l_j)}\right)\right]. \tag{B5}$$

4. Estimate the optimal values of $h_0$ and $\beta$ using the LCV criterion. Finally, the adaptive KDE to the LF at the bin $\Delta z$ is estimated as

$$\hat{\phi}_{1a}(z=z_0,L) = \hat{p}_a(z=z_0,L|h_0,\beta)n(\Omega\frac{dV}{dz})^{-1}, \tag{B6}$$

where $z_0 = (Z_1+Z_2)/2$.

### C. THE ADAPTIVE KDE CONSIDERING THE WEIGHTING

The adaptive KDE considering the weighting is

$$\hat{f}_{wa}(x,y) = \frac{1}{N_{\text{eff}}}\sum_{j=1}^{n}\frac{w_j}{\lambda_1(x_j,y_j)\lambda_2(x_j,y_j)}\left\{K\left(\frac{x-x_j}{\lambda_1(x_j,y_j)},\frac{y-y_j}{\lambda_2(x_j,y_j)}\right)+K\left(\frac{x-x_j}{\lambda_1(x_j,y_j)},\frac{y+y_j}{\lambda_2(x_j,y_j)}\right)\right\}. \tag{C7}$$

where $\lambda_1(x_j, y_j)$ and $\lambda_2(x_j, y_j)$ are calculated via Equation (16). The leave-more-out estimator is

$$\hat{f}_{\text{wa},(-i)}(x_i, y_i) = \frac{2}{(2N_{\text{eff}} - \eta_i)} \left\{ \sum_{j \in J_i} \frac{w_j K\left(\frac{x_i - x_j}{\lambda_1(x_j, y_j)}, \frac{y_i - y_j}{\lambda_2(x_j, y_j)}\right)}{\lambda_1(x_j, y_j)\lambda_2(x_j, y_j)} + \sum_{j \in J'_i} \frac{w_j K\left(\frac{x_i - x_j}{\lambda_1(x_j, y_j)}, \frac{y_i + y_j}{\lambda_2(x_j, y_j)}\right)}{\lambda_1(x_j, y_j)\lambda_2(x_j, y_j)} \right\}, \quad \text{(C8)}$$

where $J_i = \{j : x_i \neq x_j \text{ and } y_i \neq y_j, \text{for } j = 1, 2, ..., n\}$, and $J'_i = \{j : x_i \neq x_j, \text{for } j = 1, 2, ..., n\}$, for i=1,2,...n; $\eta_i$ denotes the total number of terms that are excluded from the two summations. Finally, the weighted version adaptive KDE LF is given by

$$\hat{\phi}_{\text{a}}(z, M) = \frac{N_{\text{eff}}(Z_2 - Z_1)\hat{f}_{\text{wa}}(x, y|h_1, h_2, \beta)}{(z - Z_1)(Z_2 - z)\Omega \frac{dV}{dz}}. \quad \text{(C9)}$$